\documentclass[11pt,a4paper]{article}
\usepackage[margin=1in]{geometry}
\usepackage[T1]{fontenc}
\usepackage[utf8]{inputenc}
\usepackage{amsmath,amsthm}
\usepackage{newtxtext,newtxmath}
\usepackage{microtype}
\usepackage{graphicx,multirow,mathrsfs}
\usepackage[title]{appendix}
\usepackage{xcolor,textcomp,manyfoot,booktabs,placeins,enumitem}
\usepackage[numbers,sort&compress]{natbib}
\usepackage{xurl}
\definecolor{brandburgundy}{RGB}{128,0,32}
\usepackage[colorlinks=true,linkcolor=brandburgundy,citecolor=brandburgundy,urlcolor=brandburgundy]{hyperref}
\usepackage{fancyhdr}
\graphicspath{{figures-new/}}

\theoremstyle{definition}

\makeatletter
\renewcommand\section{\@startsection{section}{1}{\z@}{-3.5ex\@plus-1ex\@minus-.2ex}{2.3ex\@plus.2ex}{\normalfont\large\bfseries\color{brandburgundy}}}
\renewcommand\subsection{\@startsection{subsection}{2}{\z@}{-3.25ex\@plus-1ex\@minus-.2ex}{1.5ex\@plus.2ex}{\normalfont\normalsize\bfseries\color{brandburgundy}}}
\renewcommand\subsubsection{\@startsection{subsubsection}{3}{\z@}{-3.25ex\@plus-1ex\@minus-.2ex}{1.5ex\@plus.2ex}{\normalfont\normalsize\bfseries\color{brandburgundy}}}
\makeatother
\hypersetup{pdftitle={Do Cryptocurrency Markets Differentiate Infrastructure from Regulatory Shocks? A Multi-Moment Event Study with Dependence-Robust Inference},pdfauthor={Murad Farzulla},pdfsubject={Author manuscript; forthcoming in Digital Finance}}
\title{\color{brandburgundy}Do Cryptocurrency Markets Differentiate Infrastructure from Regulatory Shocks? A Multi-Moment Event Study with Dependence-Robust Inference}
\author{Murad Farzulla\\[3pt]
\small King's Business School, King's College London; Dissensus\\
\small London, United Kingdom\\
\small\href{mailto:murad@dissensus.ai}{murad@dissensus.ai}}
\date{\textbf{Forthcoming, Digital Finance}\\[3pt]
\small Author manuscript in Dissensus house style, 12 September 2026}
\begin{document}
\maketitle
\begin{abstract}
Do cryptocurrency markets process infrastructure shocks (exchange outages, protocol exploits, and other infrastructure-side developments) differently from regulatory shocks (enforcement actions, policy changes)? We study the question at both moments of the return distribution over one shared sample (50 events, six assets, January 2019--August 2025), fitting an asymmetric GARCH model with exogenous regressors (GJR-GARCH-X) under a dependence-aware inference design. We lead with a methodological move: treating event inclusion as an explicit design parameter. Rather than asserting the standard selection-on-the-dependent-variable objection away, we trace the infrastructure-regulatory variance differential across alternative inclusion screens and report how sensitive the estimate is to those rules. The resulting point-estimate pattern is a descriptive scope condition, not an inferential comparison between screens. Under curated, high-salience event identification the differential is sizeable---a $3.49\times$ point-estimate multiplier---but it is selection-conditional: applying the impact filter mechanically to a broad reconstructed candidate pool yields approximately $0.5$--$1.6\times$.

Identification is only half the story; inference is the other half. The curated variance differential is not significant against its fitted sharp per-asset-equality null under the specified conditional fixed-path Student-$t$-copula bootstrap ($p\approx0.39$), while the floored recursive sensitivity gives one-sided $p=0.025$ at baseline and $0.041$ under the asset-specific high-variance-regime control, making the second-moment verdict scheme- and implementation-sensitive. Across the reported dependence inputs, the six-contrast effective sample size is $1.33$--$2.35$, and the one-sided effective-df sensitivity for the dispersion statistic gives $p=0.044$--$0.116$. A naive i.i.d.\ test treating the two sets of six per-asset coefficients as independent samples had produced an apparently decisive result, but that significance is not robust to dependence and heavy-tail corrections. At the first moment, the CAR difference is $+8.69$ percentage points and is not distinguishable from zero under the event-level block bootstrap ($p=0.202$). On this sample, the asymmetry remains directional, selection-conditional, inference-scheme-sensitive, and unresolved; the contribution is a portable diagnostic toolkit---an inference ladder and an internal Monte-Carlo calibration study---demonstrated through a self-correction of the author's earlier significance claim.
\end{abstract}


\section{Introduction}\label{sec:intro}

A recurring question in cryptocurrency risk management is whether markets differentiate between the \emph{type} of news shock, or whether shocks from different channels are processed equivalently. Two event categories are natural candidates for distinct processing. Infrastructure failures---exchange outages, protocol exploits, bridge hacks, network disruptions---impair trading and settlement mechanically, and hacking incidents are associated with sharp contemporaneous return and volatility responses \citep{ChenEtAl2023, Grobys2021}. Regulatory shocks---enforcement actions, policy changes, jurisdictional rulings---operate through expectation channels, requiring participants to reassess legal constraints and compliance costs; market valuations react to regulatory news even where trading does not systematically relocate across jurisdictions \citep{AuerClaessens2018, FeinsteinWerbach2021}. If markets price these channels differently, the difference should show up somewhere in the return distribution: in the first moment (cumulative abnormal returns), the second moment (conditional variance), or both.

An earlier version of this work answered the question affirmatively at the second moment \citep{Farzulla2025Infra}. Estimating GJR-GARCH-X models on a six-asset panel, it reported an infrastructure-event conditional-variance response $5.7\times$ as large as the regulatory-event response, with an apparently decisive significance level.\footnote{That predecessor circulates as a Research Square preprint (10.21203/rs.3.rs-8323026). The present paper corrects and supersedes its $p<0.001$ significance claim; the first-moment companion preprint (Ref.~\citealp{Farzulla2025SameReturns}) is likewise superseded by the unified analysis here.} That $5.7\times$ is the prior paper's high-salience-curated estimate ($\bar\delta_{\text{infra}}/\bar\delta_{\text{reg}}=2.385/0.419$) under its canonical estimator and rolling winsorisation; the present standalone pipeline does not reproduce it, and reports a curated point estimate of $3.49\times$ (global-clip winsorisation, canonical estimator), which is the headline figure throughout this paper (Section~\ref{sec:spec_stability}). The companion first-moment analysis \citep{Farzulla2025SameReturns} found no return-level difference, and the two were presented as a tidy story: the market differentiates shock types through the risk channel, not expected returns---structure that lives in the second moment.

That story does not survive scrutiny. This paper re-examines the same events at both moments under one dependence-aware design; at the first moment the difference was never significant, and at the second the earlier $p\approx 0.001$ proves not to be robust---the conditional inference of record does not reject, while a floored recursive-propagation sensitivity rejects one-sided but requires material positivity repair (Appendix~\ref{app:bootstrap}). The lead contribution is therefore not a settled finding about cryptocurrency but a finding about inference, told as a self-correction: an estimate the author himself previously reported as a significant fivefold effect becomes a directional, scheme- and implementation-sensitive result once cross-asset dependence and heavy tails are addressed. We think this is worth reporting plainly. The earlier significance claim depended on untenable independence assumptions; the magnitude and direction remain unresolved rather than disproved.

\subsection{What we test, and what we find}

We assemble a single shared master classification: 50 carefully identified events (26 infrastructure, 24 regulatory) across six cryptocurrencies (BTC, ETH, XRP, BNB, LTC, ADA), January 2019 to August 2025, with a decomposed GDELT sentiment series. We then test the infrastructure-regulatory difference at both moments of the return distribution using two estimand-specific procedures joined by one principle: the asset series share a largely common event calendar and are strongly cross-correlated, so per-asset or per-observation quantities cannot be treated as independent replications. Five variance-model series support all 50 event dates, while BNB supports the 45 on or after its entry.

\begin{description}[leftmargin=0pt,labelindent=0pt]
\item[First moment (returns).] An event-level block bootstrap that resamples whole events (preserving the cross-sectional correlation of returns within an event) and an event-level Welch $t$-test on event-level mean cumulative abnormal returns (CARs). On the shared six-asset/50-event basis, infrastructure CARs are $-0.64\%$ and regulatory CARs $-9.33\%$, a difference of $+8.69$ percentage points (infrastructure \emph{less} negative). The block-bootstrap two-sided $p$ is $0.202$, the event-level Welch $p$ is $0.216$, and the 95\% confidence interval $[-4.50\%, +22.22\%]$ crosses zero comfortably. The observed effect ($\approx 9$ pp, Cohen's $d=0.35$) sits well below the minimum detectable effect for this sample ($\approx 19$ pp at 80\% power)---an underpowered failure to reject, not a demonstration of equality.

\item[Second moment (variance).] A Student-$t$-copula CCC-GARCH-X parametric bootstrap. Standardised innovations are redrawn from the fitted Student-$t$ margins (degrees of freedom $\nu \approx 3$) under the cross-asset standardised-residual correlation, and the cross-asset mean coefficient difference is re-estimated on each replicate. The observed point-estimate multiplier is directional---$3.49\times$ at baseline, attenuating to $\approx 1.9$--$2.9\times$ under structural-break controls---but the conditional bootstrap returns absolute-statistic $p=0.389$ at baseline and $0.399$ under the asset-specific high-variance-regime control (one-sided $p=0.388$ and $0.394$). The multiplier is directional and not significant under this scheme; the floored recursive sensitivity is reported separately.
\end{description}

Both moments thus show directional point estimates that are not significant under the specified conditional procedures. We are deliberate about the framing. This is a \emph{dual failure to reject under those procedures}, not a claim that infrastructure and regulatory shocks are identical or that the effects are zero. At the first moment the study is plainly underpowered; at the second the point estimate is non-trivial and the conclusion changes with the resampling implementation. The honest summary is: under the specified conditional fixed-path procedures, we cannot reject equality of the first-moment group responses or the second moment's sharp per-asset-equality null; Appendix~\ref{app:bootstrap} reports a floored recursive second-moment sensitivity that rejects that sharp null under the prespecified one-sided test but requires material positivity repair.

\subsection{Why the earlier significance was not robust}

The earlier variance headline rested on two inferential moves that, examined closely, produced excessive precision. First, the significance test treated the six infrastructure coefficients and six regulatory coefficients as independent samples. They are not: five asset models contain all 50 master-calendar events and BNB contains 45, so all six overlap on 45 common events; the assets also correlate $0.54$--$0.83$ (mean $\bar\rho \approx 0.68$, Figure~\ref{fig:correlation_heatmap}), leaving an effective sample size far below six. This is textbook pseudoreplication and accounts for much of the apparent $p \approx 0.001$. Second, the cross-asset bootstrap that was meant to address dependence drew its standardised innovations from a Gaussian even though each model is fitted with Student-$t$ errors at $\nu \approx 3$. A unit-variance Student-$t$ at $\nu \approx 3$ concentrates its mass in the body and carries its variance in rare extreme tails; the Gaussian draw therefore mis-specifies the null distribution. In the fitted internal-calibration experiments this misspecification is strongly anti-conservative. Addressing the first issue---with event-level proxies and a design-effect calculation reported under an explicit effective-degrees-of-freedom sensitivity convention---places the relevant $p$-values between approximately $0.04$ and $0.69$, depending on estimand and procedure; replacing the Gaussian innovation draw in the fixed-path cross-asset bootstrap moves its one-sided $p$ from $0.113$ to $0.388$ (absolute-statistic $p$ from $0.113$ to $0.389$).

We organise this as an \emph{inference ladder} (Section~\ref{sec:artifact}, Table~\ref{tab:ladder}): seven related procedures and proxies placed side by side, with their differing estimands and dependence or tail assumptions stated explicitly. The historic headline ($t=4.768$, $p=0.0008$) sits at one end and the conditional Student-$t$-copula bootstrap (one-sided $p=0.388$) at the other; the rungs bracket the comparison rather than moving monotonically. An internal Monte-Carlo calibration study then simulates a thousand panels under the fitted fixed-design null: within that DGP, the naive i.i.d.\ rule rejects $35.1\%$ of true-null panels at the $5\%$ level, whereas the Student-$t$-copula rule rejects $4.6\%$. The load-bearing result is the severe over-rejection of the naive rule; the near-nominal copula result is a self-consistency diagnostic for the fitted calibration DGP, not independent validation of the plug-in bootstrap.

\subsection{What survives}

Two substantive results do survive the re-examination and provide the paper's ballast.

\emph{A descriptive scope pattern in the variance differential.} The directional second-moment point estimate ($3.49\times$, not significant under the conditional inference of record) is much larger under curated, high-salience event identification than under a mechanical impact screen. Applying the filter at increasing strictness to a broad reconstructed candidate pool yields multipliers of approximately $0.5$--$1.6\times$, none significant under the descriptive Welch comparison; the like-for-like two-asset screen is approximately $1.6\times$, not $3.49\times$. We treat the event-inclusion rule as an explicit researcher degree of freedom and report this inclusion-rule sensitivity, without dependence-robust inference for differences across screens. The reconstructed census also shows a higher mechanical-screen pass rate for regulatory than infrastructure candidates ($77.4\%$ versus $64.6\%$), which addresses that narrow filter mechanism without adjudicating the neutrality of upstream manual curation.

\emph{A weekly, litigation-concentrated sentiment lead.} At the constructed index's weekly frequency, Global Database of Events, Language, and Tone (GDELT) news sentiment Granger-leads volatility for 10 of 18 asset-sentiment pairs, 7 surviving a Benjamini--Hochberg false-discovery-rate (FDR) correction at $q<0.05$ (Section~\ref{sec:granger}). A prior daily test forward-filled the weekly predictor and found no sentiment-to-volatility lead, but that is a different, poorly matched frequency specification and cannot be read as evidence against a weekly association. The 7 FDR survivors remain significant under the reported Toda--Yamamoto lag augmentation, Bitcoin co-movement, and classified-event controls. The warm-up-handling diagnostic is less definitive, and the associations are concentrated in SEC-litigation assets: only 3 of 7 retain a pre-litigation lead, while XRP's three associations are not detected before SEC v.\ Ripple in this split. We therefore report a litigation-concentrated weekly lead-lag association requiring further validation, not a general anticipatory channel.

\subsection{Contribution and structure}

The contribution is threefold. First, and centrally, an inference cautionary tale: a worked, self-refuting demonstration that pseudoreplication and tail mis-specification yield severe over-rejection in the fitted calibration experiment, while the conditional fixed-path result and floored recursive sensitivity leave the substantive empirical verdict unresolved. We make this concrete with an \emph{inference ladder} and an internal Monte-Carlo calibration study. Second, a descriptive mapping of how the variance point estimate responds to the event-inclusion screen. Third, the weekly sentiment-leads-volatility result, recovered once the frequency mismatch is addressed. We are explicit about what this paper shares with, and adds to, the two earlier analyses it supersedes: it draws on the same event sample as the first-moment companion \citep{Farzulla2025SameReturns} and re-examines the same second-moment estimand as \citep{Farzulla2025Infra}, but what is new---and present in neither---is the unified single-sample multi-moment design, the dependence-aware re-analysis that overturns the earlier variance significance, and the event-inclusion sensitivity map, so this is a methodological re-analysis that changes the prior conclusion, not a partition of an existing result. Section~\ref{sec:literature} situates the work; Section~\ref{sec:methods} describes the shared sample and dependence-aware procedures; Section~\ref{sec:results_returns} reports the first moment and Section~\ref{sec:results_var} the second; Section~\ref{sec:artifact} sets out the inference ladder, the four lessons, and the Monte-Carlo study; Section~\ref{sec:robustness} reports the surviving robustness battery and the sentiment result; Sections~\ref{sec:discussion}--\ref{sec:conclusion} discuss and conclude.

% ============================================================================
% LITERATURE REVIEW
% ============================================================================

\section{Literature and Background}\label{sec:literature}

\subsection{Cryptocurrency volatility and event studies}

Cryptocurrency volatility is heavy-tailed, strongly clustered, and near-integrated. GARCH-family models are the standard tool for cryptocurrency volatility, and specification choice within the family matters \citep{Katsiampa2017, ChuEtAl2017}; asymmetric specifications are often preferred, and the asymmetry runs opposite to the equity leverage effect: both \citet{CheikhEtAl2020} and \citet{BaurDimpfl2018} find that positive return shocks amplify volatility more than negative ones. Cross-asset dependence is high (Figure~\ref{fig:correlation_heatmap}): \citet{MakarovSchoar2020} document persistent cross-exchange price gaps and a common order-flow component that explains 80\% of bitcoin returns, and \citet{KimEtAl2021} construct a market-wide cryptocurrency volatility index; the same systemic co-movement motivates aggregate crypto risk indices \citep{Farzulla2025ASRI}. This dependence is precisely what makes naive cross-asset inference unreliable, a point this paper makes central.

Event-study methodology \citep{CampbellEtAl1997, mackinlay1997} faces particular challenges in crypto: continuous 24/7 trading eliminates natural event windows, global participation creates timezone-dependent liquidity, and retail dominance changes information processing---early evidence suggests uninformed users approach cryptocurrency as an alternative investment vehicle rather than a means of payment \citep{GlaserEtAl2014}. \citet{ChenEtAl2023} examine how hacking incidents affect the price dynamics of Bitcoin spot and futures; \citet{Grobys2021} models the volatility response to hacking incidents with GARCH-type specifications and post-event dummies. The two literatures use incompatible windows---short for infrastructure (mechanical impact), long for regulation (gradual absorption)---which precludes direct comparison. The cleanest way to compare them is to hold the sample, the model, and the inference fixed and vary only the event type, which is what we do.

\subsection{Dependence-aware event-level inference}

The methodological core of this paper is the explicit treatment of dependence. When abnormal returns are cross-sectionally correlated, standard event-study $t$-tests over-reject \citep{kolari2010event}, and event-induced variance compounds the problem \citep{boehmer1991event}. The cluster-robust literature \citep{cameron2008bootstrap, petersen2009estimating, mackinnon2023cluster} recommends resampling at the level carrying the shared shock; related small-cluster procedures include the grouped $t$ approach of \citet{ibragimov2010tstatistic}. Recent work emphasises that event-study inference is fragile to specification and identification choices more broadly \citep{goldsmithpinkham2025factor, casini2025shock}. Our first-moment implementation resamples event-level cross-asset means and supplements it with an ordinary Welch test on those event-level observations; it is not the Ibragimov--M\"uller partition procedure. The second-moment implementation instead uses a joint cross-asset parametric bootstrap. Both address contemporaneous cross-asset dependence; the remaining between-event dependence from overlapping CAR windows is stated separately as a limitation.

\subsection{Sentiment and information processing}

GDELT-based news sentiment provides a public alternative to social-media measures \citep{ShenEtAl2019, PhillipsGorse2018}; \citet{AuerClaessens2018} find cryptocurrency valuations and volumes react to regulatory news, while \citet{FeinsteinWerbach2021} find no systematic cross-jurisdiction reallocation of trading activity after regulatory announcements; \citet{LiuTsyvinski2021} show that attention proxies carry return information. Closest to our infrastructure--regulatory contrast, \citet{LyocsaEtAl2020} document that regulatory announcements and exchange-hacking news each move Bitcoin volatility. To our knowledge, no prior study decomposes GDELT into thematic (infrastructure vs regulatory) components or examines whether such coverage Granger-leads volatility at the appropriate frequency---a gap our weekly Granger analysis addresses.

% ============================================================================
% DATA AND METHODOLOGY
% ============================================================================

\section{Data and Methodology}\label{sec:methods}

The design principle is one shared event sample and one dependence-aware logic, implemented with estimand-specific procedures at the two moments. We first describe the sample, then the two implementations.

\subsection{Shared Sample}\label{sec:sample}

\subsubsection{Assets and prices}

Six cryptocurrencies are selected for continuous trading history over January 2019--August 2025: Bitcoin (BTC), Ethereum (ETH), XRP, Binance Coin (BNB), Litecoin (LTC), and Cardano (ADA). CoinGecko daily closes yield 2,434 observations for BTC, ETH, XRP, LTC, and ADA and 2,191 for BNB (whose cached series begins 1 September 2019 following its chain migration; 14,361 total); continuity applies from each asset's entry. Accordingly, the five full-history variance models support all 50 event dates and BNB supports the 45 on or after its entry: the shared 50-event sample is a master classification, not identical per-asset support. Variance models use daily log returns in percentage points, winsorised by a global $0.5$/$99.5\%$ clip; the CAR engine retains arithmetic percentage returns and uses wider cached histories (beginning as early as 2013 and extending into September 2025) only to populate estimation and event windows, without expanding the event sample. Table~\ref{tab:descriptive} and Figure~\ref{fig:correlation_heatmap} report raw-return moments and correlations; second-moment inference instead uses fitted standardised-residual correlation, while first-moment inference aggregates assets within events.

\subsubsection{Event sample}

The event sample is the unified basis for both moments: 50 events, classified binary infrastructure (26) versus regulatory (24), identified by a systematic four-stage protocol. Events are dated by the UTC calendar day on which the defining information became public (filing or announcement), not by market reaction or procedural confirmation. Multi-day or bundled sequences are anchored to the decisive public event---for FTX, the 11 November 2022 Chapter~11 filing rather than the 6--8 November withdrawal-halt reports; for the March 2023 banking episode, the receivership announcement rather than the stablecoin-issuer disclosure that followed it hours later---and in both cases the $[-3,+3]$ variance window covers the alternative anchors. Announcement timestamps were verified against primary sources, which is how one dating error was caught: the Celsius withdrawal freeze was announced at 02:10~UTC on 13 June 2022 (22:10 local time on 12 June) and is dated accordingly to 13 June. Candidates were drawn from five independent sources (blockchain news aggregators, mainstream financial media, official regulatory filings, security-incident databases, and exchange status pages), screened for verifiable UTC timestamps, public documentation, and cross-sectional relevance---a $\geq 2$-asset impact criterion (strict form: at least two assets moving $|r|>1$ sample~SD within $[-1,+1]$ of the event date) applied as a curation guideline rather than a mechanical gate (Section~\ref{sec:selection_bias} audits the divergence: 13 of the curated 50 fail the strict mechanical form of the screen)---classified by primary information channel, and resolved by manual expert curation to the final 50. Infrastructure events originate in the technology and market-infrastructure layer (exchange failures and outages, protocol exploits and upgrades, network disruptions, crypto-exposed bank failures, and infrastructure-side adoption milestones); regulatory events alter the legal environment (enforcement actions, legislative developments, policy announcements, jurisdictional changes). Boundary cases (e.g.\ exchange enforcement actions, stablecoin depegs) were assigned by primary channel; the resulting split is 26 infrastructure / 24 regulatory. Selected events appear in Appendix~\ref{app:events}.

Both moments are estimated on \emph{this} 50-event classification, and every event contributes a valid event-level first-moment measurement, so the event counts remain 26 versus 24. The unified CAR gate requires at least 120 estimation-window observations and a complete 36-observation $[-5,+30]$ event window; it uses 293 of the nominal 300 asset-event cells. BNB is unavailable for the six earliest events (five with no price history and a sixth below the 120-observation minimum), and its cell for the final event (7 August 2025) is omitted because the BNB source series ends on 31 August, before the $+30$ horizon. Each affected event is averaged over the remaining five assets, so no event is dropped. We note that the first-moment estimates are run on the full binary 50-event split (mixing positive- and negative-valence events) to match the variance basis exactly; a legacy adverse-event analysis is retained for historical comparison only (Section~\ref{sec:results_returns}), but it uses a different event snapshot, dating, and four-asset universe and therefore is not a valence control for the unified 50-event estimates.

\subsubsection{Sentiment}

GDELT-based sentiment indices \citep{LeetaruSchrodt2013} are constructed by a three-stage decomposition: hierarchical keyword filtering separating regulatory content (``SEC,'' ``regulation,'' ``compliance'') from infrastructure content (``hack,'' ``exploit,'' ``outage''); volume-weighted tone aggregation with recursive $z$-score detrending over 52-week rolling windows; and thematic decomposition weighted by article coverage,
\begin{align}
S_t^{\text{REG}} &= S_t^{\text{GDELT}} \times \text{Proportion}_t^{\text{REG}}, \\
S_t^{\text{INFRA}} &= S_t^{\text{GDELT}} \times \text{Proportion}_t^{\text{INFRA}}.
\end{align}
Strictly, the thematic components are \emph{coverage-weighted aggregate tone}---the aggregate index scaled by each theme's article share---rather than within-theme tone, so their thematic interpretation is correspondingly approximate. The constructed index is weekly (GDELT itself updates continuously; the weekly aggregation is a construction choice governed by coverage sparsity in the filtered crypto-news subset, and a daily reconstruction from raw GDELT is left to future work), with up to a 7-day temporal mismatch to daily volatility. Its 52-week rolling $z$-score uses a 26-observation minimum, leaving the first 25 weekly values (about $7\%$ of the 345-week series) undefined; the daily variance-equation pipeline zero-fills those warm-up values. The implementation constructs that daily control on the six-asset common calendar (beginning 2 September 2019) before reindexing to each asset, so the five longer histories also receive zeros for 24 June--1 September 2019 (70 days; BNB is unaffected), including the August 2019 Litecoin event window. This is a nuisance-control alignment limitation: an audit-only union-calendar observed refit leaves the multiplier at $3.49\times$ to reported precision, but the bootstrap was not recalibrated and no separate inferential claim is drawn. For that equation the weekly index is then forward-filled to daily, entering as a within-week step function---adequate for a slow-moving level control, though it carries no within-week timing information; this limitation is exactly why the lead--lag analysis of Section~\ref{sec:granger} is run at the native weekly frequency rather than on the forward-filled daily series. The native weekly frequency is decisive for the Granger analysis of Section~\ref{sec:granger}.

\subsection{First Moment: Event-Level Block Bootstrap}\label{sec:methods_returns}

The first-moment instantiation follows standard event-study methodology \citep{mackinlay1997, kothari2007econometrics} with cryptocurrency adaptations and an event-level, dependence-aware inference layer.

\textbf{Windows.} The predecessor engine uses a 250-calendar-day span with inclusive endpoints, from event day $-281$ through event day $-31$ (251 daily observations when complete; 120 minimum for newer assets). Days $-30$ through $-1$ are outside estimation, while days $-5$ through $-1$ also form the pre-event segment of the $[-5,+30]$ CAR window. The unified gate admits an asset-event cell only when all 36 daily CAR-window observations are present (shorter $[0,+1]$ to $[0,+5]$ windows were examined as robustness in the predecessor returns analysis).

\textbf{Overlapping windows.} Same-type $[-3,+3]$ variance indicators take the union of overlapping windows, so a day is not double-counted within a type; opposite-type overlap activates both indicators, so the shared day is not assigned arbitrarily and the GJR-GARCH-X estimates both contributions jointly. Each $[-5,+30]$ CAR is computed separately, so a shared calendar day can enter two event CARs; the event-level bootstrap preserves within-event cross-asset dependence but treats event IDs as independent and does not jointly resample this overlap-induced between-event dependence. The direction of that simplification is not determined a priori: shared days between opposite-type events co-move the group means and can partly cancel in their difference. Its extent is modest: 24 of the $\binom{50}{2}=1{,}225$ event pairs have overlapping CAR windows (mean $16.5$ shared days), and only 3 pairs overlap in the variance windows.

\textbf{Abnormal returns.} Suppressing the event index $e$, the primary specification is the constant-mean model, $AR_{it} = R_{it} - \bar{R}_i$, with $\bar{R}_i$ estimated over that event's estimation window; a market model using BTC as proxy, $AR_{it} = R_{it} - (\hat\alpha_i + \hat\beta_i R_{\text{BTC},t})$, is the robustness alternative. Cumulative abnormal returns over $[\tau_1,\tau_2]$ are $CAR_i(\tau_1,\tau_2) = \sum_{t=\tau_1}^{\tau_2} AR_{it}$.

\textbf{The inference problem.} Pooling asset-event observations gives a nominal $50\times6=300$ observations (293 usable; Section~\ref{sec:sample}), but the event is the resampling unit and assets within it are strongly correlated. An i.i.d.\ $t$-test would therefore overstate power---the first-moment analogue of the variance-side pseudoreplication.

\textbf{The dependence-aware solution.} We average usable asset CARs within each event (equal-weighting events whether five or six assets contribute), then resample infrastructure and regulatory event means separately at their original group sizes, retaining cross-sectional dependence in the aggregation \citep{cameron2008bootstrap, petersen2009estimating, boehmer1991event}. We use 5,000 seeded replications and percentile intervals. For an observed positive difference, the reported two-sided $p$ is $\min\{1,2\Pr^*(\Delta^*\leq0)\}$ (inequality reversed if negative): a group-wise bootstrap sign probability, not a pooled-label randomisation test or equality-imposed bootstrap. An event-level Welch $t$-test is the parametric cross-check. Both calculations assume independence across event IDs, subject to the overlap limitation above; together they form the specified first-moment inference.

\subsection{Second Moment: Student-\texorpdfstring{$t$}{t}-Copula CCC-GARCH-X Bootstrap}\label{sec:methods_var}

The second-moment instantiation estimates the conditional-variance response to each event type within a GJR-GARCH-X framework, with inference from a dependence-preserving parametric bootstrap calibrated to the same cross-asset correlation.

\textbf{The model.} The mean equation is a constant conditional mean, $r_t = \mu + \varepsilon_t$, justified empirically: Ljung--Box tests find serial correlation in only 8 of 18 (asset $\times$ lag) cells, the fitted AR(1) coefficient is tiny and negative everywhere ($-0.025$ to $-0.067$), and refitting with an AR(1) mean leaves the event coefficients essentially unchanged (Section~\ref{sec:robustness}). The variance equation is the GJR-GARCH-X of \citet{GlostenEtAl1993} with exogenous regressors,
\begin{equation}
\begin{split}
\sigma^2_t = \omega + \alpha \varepsilon^2_{t-1} + \gamma \varepsilon^2_{t-1}\,\mathbb{I}(\varepsilon_{t-1}<0) + \beta \sigma^2_{t-1} &+ \delta_{\text{infra}} D^{\text{infra}}_t + \delta_{\text{reg}} D^{\text{reg}}_t \\
&+ \theta_0 S^{\text{GDELT}}_t + \theta_1 S^{\text{REG}}_t + \theta_2 S^{\text{INFRA}}_t,
\end{split}
\end{equation}
where $D^{\cdot}_t$ are event dummies over the $[-3,+3]$-day window and the coefficients $\delta$ are additions to conditional variance in squared percentage points. The three sentiment terms enter jointly: $\theta_0$ loads on the aggregate normalised GDELT index $S^{\text{GDELT}}_t$, controlling for the overall sentiment level, while $\theta_1$ and $\theta_2$ load on its regulatory- and infrastructure-specific decompositions, so they capture theme-specific sentiment beyond the aggregate. The variance equation carries five exogenous coefficients; with $\omega,\alpha,\gamma,\beta,\nu$ and the sample mean $\mu$ estimated before and held fixed during variance optimisation, the information criteria in Table~\ref{tab:model_comparison} count eleven fitted quantities for GJR-GARCH-X (and 5/6 for GARCH/GJR-GARCH). Because the dummy stays on across the seven-day window, a $\delta$ shock accumulates through the $\beta\sigma^2_{t-1}$ feedback rather than producing a flat one-day shift---the within-window peak contribution is approximately $\delta(1-\beta^{k})/(1-\beta)$ after $k$ days---but this amplification is common to both event types and largely cancels in the infrastructure--regulatory ratio (the cross-asset multiplier moves only from $3.49\times$ to $\approx 3.7\times$ under full compounding), so it scales absolute magnitudes without materially affecting the headline ratio. Variance and distribution parameters are estimated by conditional maximum likelihood under Student-$t$ innovations after demeaning each asset at its sample mean.\footnote{The consistency results cited next use a Gaussian QMLE criterion; our inference of record does not use its standard errors. The replication code contains the transparent manual estimator used for the specification sweep and \texttt{FastTARCHX}, its accelerated counterpart for repeated fits and diagnostics. On the corrected baseline, their negative log-likelihoods differ by at most $1.1\times10^{-5}$, event coefficients by at most $0.0015$ squared percentage points, and the cross-asset multiplier is identical at the reported precision.}

Two near-integration concerns arise and are addressed by distinct results. First, the exogenous covariates may be persistent: \citet{HanKristensen2014} provide Gaussian-QMLE theory for GARCH-X with stationary and nonstationary covariates. This provides relevant asymptotic support for allowing persistent sentiment regressors, but it is not a direct theorem for the present Student-$t$, multi-regressor specification. Second, and separately, the GARCH \emph{volatility persistence} itself is near-integrated ($\alpha+\beta\to 1$). \citet{Nelson1990} characterises stationarity and persistence in GARCH(1,1), while \citet{Lumsdaine1996} establishes consistency and asymptotic normality of the QMLE for IGARCH(1,1) and covariance-stationary GARCH(1,1). Those asymptotic results do not guarantee reliable finite-sample Hessian standard errors at a near-boundary estimate. The Student-$t$-copula bootstrap avoids reliance on Hessian standard errors, but it remains conditional on an estimated, near-integrated DGP; it therefore addresses the analytic-standard-error problem without eliminating all near-integration sensitivity. Appendix~\ref{app:bootstrap} reports the consequential fixed-path versus recursive-path sensitivity explicitly. Covariance stationarity is enforced via the conservative bound $\alpha+\beta+|\gamma|/2\le 0.999$ (a sufficient condition: the standard symmetric-innovation condition is $\alpha+\beta+\gamma/2<1$, so the absolute-value form is strictly more restrictive when $\gamma<0$, as fitted here).

\textbf{The reported point summary} is the cross-asset infrastructure-regulatory multiplier, $\bar\delta_{\text{infra}}/\bar\delta_{\text{reg}}$. Because a ratio is unstable when the regulatory denominator approaches zero, formal bootstrap inference uses the signed difference $\bar d=\bar\delta_{\text{infra}}-\bar\delta_{\text{reg}}$; the implemented null is the sharp asset-by-asset restriction that replaces the two type coefficients by one common event coefficient within each asset (the common coefficients remain asset-specific). This restriction implies $\bar d=0$ but is stronger than the composite mean null: the resulting $p$-value is not calibrated for every heterogeneous configuration whose positive and negative asset contrasts merely average to zero. The multiplier is re-estimated and reported on each specification but is not the bootstrap test statistic.

\textbf{The inference problem.} Within each event type, the six per-asset coefficients are not independent: five assets use all 50 master-calendar event dates, BNB uses 45, all six overlap on those 45 dates, and their returns correlate $\bar\rho \approx 0.69$. A test treating the coefficients as six i.i.d.\ observations is pseudoreplication and overstates precision (Section~\ref{sec:artifact}). The relevant innovation distribution is also heavy-tailed---the fitted Student-$t$ degrees of freedom are $\nu \approx 3$---so a bootstrap that draws Gaussian innovations mis-specifies the null.

\textbf{The dependence-aware solution.} We use a Student-$t$-copula CCC-GARCH-X bootstrap ($B=2000$): latent multivariate-normal draws at fitted $R_z$ receive shared chi-square mixing, map through uniforms to the fitted per-asset Student-$t$ margins, and are rescaled to unit variance, using fitted $R_z$ as the latent dependence input while preserving copula tail dependence. Each draw jointly simulates the asset panel under one common event-type coefficient per asset, refits all six models unrestricted, and recomputes $\bar d=\frac{1}{6}\sum_i(\hat\delta_{\mathrm{infra},i}-\hat\delta_{\mathrm{reg},i})$; the observed multiplier is not the bootstrap statistic. The fixed-path scheme is the conditional inference of record and the recursive scheme a sensitivity (Appendix~\ref{app:bootstrap}). A failed or $|\delta|>50$ refit receives one deterministic six-start rescue before exclusion. After rescue, the baseline Student-$t$ bootstrap retains all $2000/2000$ draws; its Gaussian comparator retains $1999/2000$. The procedure in this paper remains conditional on the fitted path and model.

\textbf{One principle, two procedures.} Both moments represent observed cross-asset dependence rather than treating correlated quantities as independent: the first resamples event-level cross-asset means nonparametrically, while the second simulates the joint heavy-tailed innovation structure parametrically. Their assumptions and remaining dependence limitations differ as stated above.

\subsection{Model selection and multiple testing}

For the variance models, three nested specifications are compared per asset (GARCH, GJR-GARCH, GJR-GARCH-X) by AIC; Benjamini--Hochberg corrections \citep{BenjaminiHochberg1995} control the false discovery rate across the 18-pair weekly Granger family (Section~\ref{sec:granger}). For the structural-break analysis, change-points are detected by the Pruned Exact Linear Time (PELT) algorithm \citep{KillickFearnheadEckley2012} with Bai-Perron-style trimming \citep{BaiPerron1998, BaiPerron2003}.

% ============================================================================
% RESULTS — FIRST MOMENT
% ============================================================================

\section{Results: First Moment (Returns)}\label{sec:results_returns}

\subsection{Descriptive statistics}

\begin{figure}[htbp]
\centering
\includegraphics[width=0.78\linewidth]{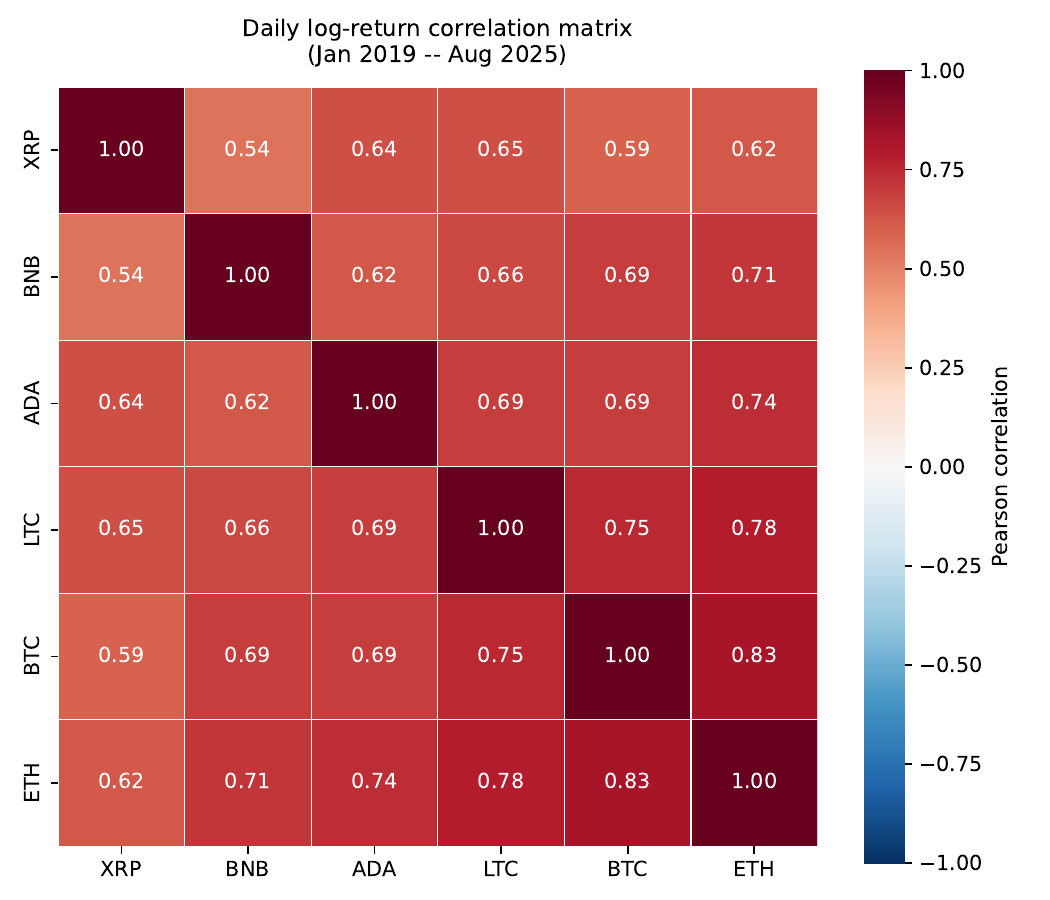}
\caption{Raw (unwinsorised) daily log-return correlation matrix across the six-asset panel, January 2019--August 2025. BTC and ETH form the most tightly correlated pair (0.83); XRP correlates least with the others (0.54--0.65). The mean pairwise correlation is $\bar\rho \approx 0.68$; the bootstrap itself uses the fitted standardised-residual correlation matrix.}\label{fig:correlation_heatmap}
\end{figure}

The panel comprises 2,434 daily observations per asset (2,191 for BNB, which lists later; 14,361 total) (Table~\ref{tab:descriptive}). The heavy tails---excess kurtosis from $9.7$ to $29.4$, far above the Gaussian benchmark---motivate Student-$t$ innovations for the variance models, while skewness is mild and mixed in sign across assets; the high cross-asset correlation (Figure~\ref{fig:correlation_heatmap}) is the inference-relevant feature.

\begin{table}[htbp]
\centering
\caption{Descriptive Statistics: Daily Returns (\%). Moments are computed on raw daily log returns; returns are winsorised (global-clip, $0.5$/$99.5\%$) for the GARCH-X estimation only. BNB lists later, giving a shorter series. Annualised volatility uses the conventional $\sqrt{252}$ scaling for comparability and is illustrative only.}\label{tab:descriptive}
\small
\begin{tabular}{@{}lrrrrrr@{}}
\toprule
\textbf{Stat} & \textbf{BTC} & \textbf{ETH} & \textbf{XRP} & \textbf{BNB} & \textbf{LTC} & \textbf{ADA} \\
\midrule
N & 2{,}434 & 2{,}434 & 2{,}434 & 2{,}191 & 2{,}434 & 2{,}434 \\
Mean & 0.14 & 0.14 & 0.09 & 0.17 & 0.05 & 0.12 \\
Std Dev & 3.31 & 4.37 & 5.18 & 4.43 & 4.77 & 5.11 \\
Skewness & $-0.91$ & $-1.05$ & $+0.51$ & $-0.17$ & $-0.70$ & $+0.24$ \\
Excess Kurt. & 15.20 & 14.84 & 20.36 & 29.38 & 9.72 & 12.37 \\
Ann. Vol & 52.5 & 69.3 & 82.3 & 70.4 & 75.7 & 81.2 \\
\bottomrule
\end{tabular}
\end{table}

\subsection{The first-moment null}

Table~\ref{tab:returns_main} reports the infrastructure-regulatory CAR comparison on the shared six-asset/50-event basis, under both the block bootstrap and the event-level Welch $t$-test. Infrastructure CARs average $-0.64\%$ and regulatory CARs $-9.33\%$, a difference of $+8.69$ percentage points: infrastructure returns are less negative than regulatory returns. The block-bootstrap two-sided $p$ is $0.202$ (one-sided $0.101$), the event-level Welch $p$ is $0.216$, and the 95\% confidence intervals---$[-4.50\%, +22.22\%]$ (block) and $[-5.26\%, +22.65\%]$ (event-level Welch)---cross zero with room to spare. We cannot reject equality of infrastructure and regulatory return responses under either calculation.

\begin{table}[htbp]
\centering
\caption{First-moment comparison: infrastructure vs regulatory cumulative abnormal returns, shared six-asset/50-event basis. Block bootstrap resamples whole events ($B=5{,}000$); Welch $p$ is an ordinary Welch $t$-test on the event-level mean CARs. The five-asset (ex-XRP) row is a robustness check; the negative-only row is a non-comparable legacy benchmark retained for historical transparency.}\label{tab:returns_main}
\footnotesize\setlength{\tabcolsep}{4pt}
\begin{tabular}{@{}lrrrrr@{}}
\toprule
\textbf{Specification} & \textbf{CAR$_{\text{infra}}$} & \textbf{CAR$_{\text{reg}}$} & \textbf{Diff (pp)} & \textbf{Block $p$} & \textbf{Welch $p$} \\
\midrule
Six-asset (incl.\ XRP), $n=26/24$ & $-0.64\%$ & $-9.33\%$ & $+8.69$ & $0.202$ & $0.216$ \\
Five-asset (ex-XRP), $n=26/24$ & $+0.44\%$ & $-7.08\%$ & $+7.52$ & $0.298$ & $0.312$ \\
Legacy adverse-event benchmark, $n=8/7$ & $-7.9\%$ & $-9.4\%$ & $+1.5$ & $0.916$ & $0.927$ \\
\bottomrule
\multicolumn{6}{@{}p{0.97\linewidth}@{}}{\footnotesize Six-asset 95\% CI: $[-4.50\%, +22.22\%]$ (block), $[-5.26\%, +22.65\%]$ (event-level Welch). Observed Cohen's $d=0.35$; pooled SD $24.6$ pp; minimum detectable effect $\approx 19.5$ pp at 80\% power, $\alpha=0.05$. The negative-only row is the legacy first-moment basis carried over for exact comparability: it uses a different four-asset universe (BTC, ETH, SOL, ADA), the predecessor event snapshot and dating (including its historical local-date Celsius coding), and only adverse-valence events. Of the 50 binary events, 17 ($34\%$) are negative-valence (10 infrastructure, 7 regulatory); 15 (8 infrastructure, 7 regulatory) yield valid CARs on that legacy basis. The unified rows use the corrected UTC event dates, including Celsius on 13 June 2022. The legacy row's CAR magnitudes are therefore not directly comparable to the full binary 50-event rows.}
\end{tabular}
\end{table}

\textbf{First-moment sensitivity and legacy comparison.} Dropping XRP (the asset with the distinct regulatory trajectory) gives a $+7.52$-pp difference, block $p=0.298$, and event-level Welch $p=0.312$. A legacy negative-only analysis (eight infrastructure versus seven regulatory events with valid CARs on the four-asset BTC/ETH/SOL/ADA basis; see the table note) gives a $+1.5$-pp difference (block $p=0.916$, Welch $p=0.927$). Its exact two-sided event-label permutation test enumerates all $\binom{15}{8}=6{,}435$ assignments, of which 5,961 are at least as extreme, giving $p=0.926$. Under the packaged hybrid BTC-proxy specification, using the same event-equal aggregation, the negative-subset point estimate is $\Delta_{\text{MM}}=+5.88$ pp (block $p=0.524$, Welch $p=0.583$, exact-permutation $p=0.574$) and remains statistically indistinguishable from zero. The BTC leg retains constant-mean CARs because an asset cannot be regressed on itself, while the ETH, SOL, and ADA legs use event-specific BTC-market regressions; BTC is the only available proxy and is itself exposed to the events. These implemented variants all fail to reject, but they do not exhaust possible return models or dependence structures. Because event membership, dates, and asset universe change simultaneously, the difference from $+8.69$ pp cannot be attributed to valence, and this row is not a robustness test of the unified specification. Both moment-specific contrasts are positive in their own units, but higher CAR and higher conditional variance have different economic meanings; their shared algebraic sign is only descriptive context.

\subsection{This is an underpowered failure to reject, not ``no effect''}

We are careful not to over-read the null. The observed six-asset difference of $+8.69$ pp is not negligible---it corresponds to Cohen's $d=0.35$---but with a pooled standard deviation of $24.6$ pp and an event-level sample of $26$ versus $24$, the minimum detectable effect at $80\%$ power is approximately $19.5$ pp (ex-XRP, $\approx 20.6$ pp). The observed effect is under half of what this sample could resolve. The honest statement is therefore ``no statistically significant difference; wide confidence intervals; underpowered,'' not ``infrastructure and regulatory returns are identical.'' The directional point estimate (infrastructure less negative) is worth reporting; the sample simply cannot determine whether it reflects a genuine difference or sampling noise. This caveat is symmetric with the second-moment finding, to which we now turn.

% ============================================================================
% RESULTS — SECOND MOMENT
% ============================================================================

\section{Results: Second Moment (Conditional Variance)}\label{sec:results_var}

\subsection{Model estimates}

Table~\ref{tab:model_comparison} reports the three nested variance specifications. GJR-GARCH-X attains the lowest AIC for five of six assets under the controlled re-estimation on the corrected census, with BNB effectively tied (GARCH(1,1) is marginally preferred there, by $0.66$ AIC; see the table note). Persistence is near-integrated (full-sample GARCH(1,1) $\alpha+\beta$ between $0.954$ and $0.998$, mean $0.988$), consistent with the near-unit persistence documented for cryptocurrencies---integrated GARCH is the best-fitting specification for five of seven coins in \citet{ChuEtAl2017}; the structural-break analysis of Section~\ref{sec:robustness} shows somewhat lower within-regime point estimates descriptively. The GJR leverage parameter $\gamma$ is uniformly small and non-positive ($-0.034$ to $\approx 0$; strictly negative for five of six assets, with BTC's estimate numerically zero), consistent with the inverted crypto leverage effect \citep{BaurDimpfl2018} rather than with conventional equity-style asymmetry, and present already---at smaller magnitude---in the event-free GJR-GARCH fits.

\begin{table}[htbp]
\centering
\caption{Model Comparison: GARCH vs GJR-GARCH vs GJR-GARCH-X (AIC; lowest in bold)}\label{tab:model_comparison}
\small
\begin{tabular}{@{}llrrr@{}}
\toprule
\textbf{Asset} & \textbf{Model} & \textbf{AIC} & \textbf{BIC} & \textbf{LogLik} \\
\midrule
BTC & GARCH(1,1) & 11865 & 11894 & $-5927$ \\
    & GJR-GARCH & 11867 & 11901 & $-5927$ \\
    & GJR-GARCH-X & \textbf{11863} & 11927 & $-5920$ \\
\midrule
ETH & GARCH(1,1) & 13297 & 13326 & $-6643$ \\
    & GJR-GARCH & 13299 & 13333 & $-6643$ \\
    & GJR-GARCH-X & \textbf{13286} & 13350 & $-6632$ \\
\midrule
XRP & GARCH(1,1) & 13271 & 13300 & $-6631$ \\
    & GJR-GARCH & 13273 & 13308 & $-6630$ \\
    & GJR-GARCH-X & \textbf{13264} & 13328 & $-6621$ \\
\midrule
BNB & GARCH(1,1) & \textbf{11346} & 11374 & $-5668$ \\
    & GJR-GARCH & 11347 & 11382 & $-5668$ \\
    & GJR-GARCH-X & 11346 & 11409 & $-5662$ \\
\midrule
LTC & GARCH(1,1) & 13727 & 13755 & $-6858$ \\
    & GJR-GARCH & 13724 & 13759 & $-6856$ \\
    & GJR-GARCH-X & \textbf{13722} & 13786 & $-6850$ \\
\midrule
ADA & GARCH(1,1) & 14040 & 14069 & $-7015$ \\
    & GJR-GARCH & 14042 & 14076 & $-7015$ \\
    & GJR-GARCH-X & \textbf{14038} & 14101 & $-7008$ \\
\bottomrule
\multicolumn{5}{@{}p{0.78\linewidth}@{}}{\footnotesize Information criteria are from a controlled multistart re-estimation of all three nested specifications on the corrected census. The sample mean is estimated from the data and then held fixed during variance optimisation, so the table counts it: $k=5/6/11$ for GARCH/GJR-GARCH/GJR-GARCH-X. Omitting that common mean count shifts every AIC and BIC level by the same within-asset amount and leaves all comparisons unchanged. GJR-GARCH-X attains the lowest AIC for five of six assets; for BNB, GARCH(1,1) is lower by $0.66$ AIC (both round to $11346$)---effectively tied, and we do not lean on it. BIC, which penalises the five exogenous coefficients more heavily, prefers GARCH(1,1) throughout; the paper makes no model-selection claim from BIC. Reproducible via \texttt{code/c18\_model\_comparison\_refit.py}; \texttt{code/c17\_model\_comparison.py} renders the compact table. The canonical per-asset GJR-GARCH-X estimates of record are the \texttt{c14} verification run reported in Table~\ref{tab:garch_params}.}
\end{tabular}
\end{table}

For completeness, Table~\ref{tab:garch_params} reports the per-asset variance-equation vector of the selected GJR-GARCH-X model---$\omega,\alpha,\gamma,\beta$, the Student-$t$ degrees of freedom $\nu$, the persistence $\alpha+\beta+|\gamma|/2$, and the two event coefficients $\delta_{\text{infra}},\delta_{\text{reg}}$ whose cross-asset means give the curated $3.49\times$ multiplier. (The sentiment terms are retained as level controls; their coefficients are not tabulated. The sentiment--volatility relationship itself is analysed at its native weekly frequency in Section~\ref{sec:granger}.)

\begin{table}[htbp]
\centering
\caption{Per-asset baseline GJR-GARCH-X variance-equation estimates (curated 50-event sample, global-clip winsorisation). Persistence is $\alpha+\beta+|\gamma|/2\le 0.999$; parameter entries are rounded to three decimals, so recomputing persistence from the displayed values can appear to exceed the bound (e.g.\ LTC, $0.099+0.884+0.034/2=1.000$), while at full precision the constraint holds exactly ($0.99900$ for the five bound assets). The cross-asset means $\bar\delta_{\text{infra}}=2.044$ and $\bar\delta_{\text{reg}}=0.586$ give the headline $3.49\times$ multiplier. Reproducible via \texttt{code/c14\_garch\_diagnostics.py}.}\label{tab:garch_params}
\small
\begin{tabular}{@{}lrrrrrrrr@{}}
\toprule
\textbf{Asset} & \boldmath$\omega$ & \boldmath$\alpha$ & \boldmath$\gamma$ & \boldmath$\beta$ & \boldmath$\nu$ & \textbf{Persist.} & \boldmath$\delta_{\text{infra}}$ & \boldmath$\delta_{\text{reg}}$ \\
\midrule
BTC & 0.122 & 0.074 & $-0.000$ & 0.925 & 3.13 & 0.999 & 0.907 & 0.301 \\
ETH & 0.151 & 0.070 & $-0.009$ & 0.925 & 3.77 & 0.999 & 2.110 & 0.381 \\
XRP & 1.381 & 0.221 & $-0.033$ & 0.762 & 3.16 & 0.999 & 2.866 & 1.737 \\
BNB & 0.293 & 0.143 & $-0.012$ & 0.850 & 4.22 & 0.999 & 1.231 & 0.359 \\
LTC & 0.694 & 0.099 & $-0.034$ & 0.884 & 3.94 & 0.999 & 2.316 & 0.336 \\
ADA & 1.443 & 0.163 & $-0.027$ & 0.795 & 4.57 & 0.971 & 2.836 & 0.404 \\
\midrule
Mean & --- & --- & --- & --- & --- & --- & \textbf{2.044} & \textbf{0.586} \\
\bottomrule
\multicolumn{9}{@{}p{0.98\linewidth}@{}}{\footnotesize Boldface in the body of Tables 4, 6, 8--12 and 15 identifies the principal summary or focal values discussed in the text; it does not denote statistical significance.}
\end{tabular}
\end{table}

\subsubsection{Variance-equation adequacy: no residual ARCH}\label{sec:garch_diagnostics}

Before interpreting the event coefficients, we test whether the fitted GJR-GARCH-X standardised residuals retain detectable ARCH---otherwise unmodelled volatility clustering, concentrated in the 2022--2023 window where infrastructure events cluster, could be absorbed by the $D^{\text{infra}}$ dummy. Table~\ref{tab:garch_diagnostics} reports Ljung--Box $Q$ on squared standardised residuals $z_t^2$ at lags 5/10/20 and the Engle ARCH-LM test, per asset. For transparency, the portmanteau is shown under both the conventional reference $\chi^2_{\text{lag}}$ distribution and a more rejection-prone $\chi^2_{\text{lag}-2}$ degrees-of-freedom sensitivity that subtracts the ARCH and GARCH lag orders. \citet{LiMak1994} motivate model-aware diagnostics for squared residual autocorrelations, but their general asymptotic correction is not reduced here to a universal degrees-of-freedom subtraction. No asset shows significant residual ARCH at the $5\%$ level under either reported convention (the closest cell, Bitcoin's lag-5 lag-minus-two sensitivity, is $p=0.084$, while its ARCH-LM(5) is $p=0.240$). Mean$(z)\approx0$ throughout; residual excess kurtosis ($2.2$--$5.8$) is compatible with the fitted Student-$t$ margins ($\nu\approx3.1$--$4.6$). These diagnostics find no evidence of remaining linear ARCH at the tested lags. They do not prove the variance model is complete or rule out higher-order nonlinearity, regime structure, or other dynamic misspecification; those limitations prevent using a non-rejection here to certify the $3.49\times$ point estimate.

\begin{table}[htbp]
\centering
\caption{Variance-equation diagnostics on the baseline GJR-GARCH-X fit: Ljung--Box $Q$ on squared standardised residuals (reported here using the more rejection-prone $\text{df}=\text{lag}-2$ sensitivity) and Engle ARCH-LM $p$. No asset shows significant residual ARCH at $5\%$ on any reported convention (0/6 throughout).}\label{tab:garch_diagnostics}
\small
\begin{tabular}{@{}lrrrrr@{}}
\toprule
\textbf{Asset} & \textbf{LB$(z^2)$ $p_5$} & \textbf{LB$(z^2)$ $p_{10}$} & \textbf{LB$(z^2)$ $p_{20}$} & \textbf{ARCH-LM $p_5$} & \textbf{ARCH-LM $p_{10}$} \\
\midrule
BTC & 0.084 & 0.159 & 0.148 & 0.240 & 0.292 \\
ETH & 0.737 & 0.422 & 0.557 & 0.942 & 0.479 \\
XRP & 0.463 & 0.763 & 0.929 & 0.762 & 0.886 \\
BNB & 0.296 & 0.269 & 0.381 & 0.604 & 0.467 \\
LTC & 0.271 & 0.529 & 0.700 & 0.567 & 0.716 \\
ADA & 0.243 & 0.365 & 0.510 & 0.513 & 0.529 \\
\bottomrule
\multicolumn{6}{@{}p{0.92\linewidth}@{}}{\footnotesize Significant flags at $5\%$: 0/6 assets on every test. The displayed portmanteau sensitivity subtracts the ARCH and GARCH lag orders ($p+q=2$); conventional $\text{df}=\text{lag}$ values in the replication output give the same conclusion. This is a diagnostic sensitivity, not a claim that a universal Li--Mak correction equals $\text{lag}-2$.}
\end{tabular}
\end{table}

The per-asset event coefficients (Table~\ref{tab:event_coefficients}) point uniformly in the same direction---infrastructure larger than regulatory for all six assets---with a cross-asset mean of $\bar\delta_{\text{infra}}\approx 2.04$ versus $\bar\delta_{\text{reg}}\approx 0.59$ (squared percentage points). The point-estimate multiplier is $3.49\times$. The question is whether that positive cross-asset difference is statistically distinguishable from zero once the cross-asset dependence and heavy tails are respected.

\begin{table}[htbp]
\centering
\caption{Per-asset baseline event-impact coefficients (squared \% points). All six assets show infrastructure $>$ regulatory; the direction is uniform, but the multiplier varies by asset. The cross-asset mean multiplier is $3.49\times$, consistent with the baseline of Table~\ref{tab:tcopula}.}\label{tab:event_coefficients}
\small
\begin{tabular}{@{}lrrr@{}}
\toprule
\textbf{Asset} & \textbf{Infra} & \textbf{Reg} & \textbf{Ratio} \\
\midrule
ADA & 2.84 & 0.40 & 7.0$\times$ \\
XRP & 2.87 & 1.74 & 1.7$\times$ \\
LTC & 2.32 & 0.34 & 6.9$\times$ \\
ETH & 2.11 & 0.38 & 5.5$\times$ \\
BNB & 1.23 & 0.36 & 3.4$\times$ \\
BTC & 0.91 & 0.30 & 3.0$\times$ \\
\midrule
\textbf{Mean} & \textbf{2.04} & \textbf{0.59} & \textbf{3.49$\times$} \\
\bottomrule
\multicolumn{4}{@{}p{0.78\linewidth}@{}}{\footnotesize Ratios are computed from full-precision coefficients. Per-asset significance tests treat cross-correlated assets as independent and are not used for final inference. The $3.49\times$ ratio-of-means is the reported point summary, while the copula bootstrap tests the more stable signed difference $\bar\delta_{\text{infra}}-\bar\delta_{\text{reg}}$. Per-asset ratios are least stable where the regulatory coefficient is small: ADA ($0.40$, ratio $7.0\times$) and LTC ($0.34$, ratio $6.9\times$) are better read alongside their variance differences ($2.43$ and $1.98$ squared percentage points).}
\end{tabular}
\end{table}

\subsection{The second-moment null}

Table~\ref{tab:tcopula} reports the Student-$t$-copula CCC-GARCH-X fixed-path bootstrap, the specified primary inference for the second moment. At baseline the multiplier is $3.49\times$, but the absolute-statistic bootstrap $p$ is $0.389$: not significant at the 10\% level under this scheme. Adding one asset-specific high-variance-regime control gives $2.90\times$ with $p=0.399$; under the full set of regime dummies the multiplier is $1.94\times$ with $p=0.392$. Across the three fixed-path specifications the variance differential is directional ($\approx 1.9$--$3.5\times$, infrastructure larger) but not significant. The floored recursive result is reported separately and prevents treating this as a scheme-invariant conclusion.

\begin{table}[htbp]
\centering
\caption{Second-moment comparison: Student-$t$-copula CCC-GARCH-X bootstrap ($B=2000$, curated $n=50$ sample). Innovations are redrawn from the fitted Student-$t$ margins ($\nu\approx 3$) under the cross-asset standardised-residual correlation. This is the inference of record. The point estimate is directional; none of the three specifications is significant.}\label{tab:tcopula}
\small
\begin{tabular}{@{}lcccc@{}}
\toprule
\textbf{Specification} & \boldmath$\bar\delta_{\text{infra}}$ & \boldmath$\bar\delta_{\text{reg}}$ & \textbf{Multiplier} & \textbf{Absolute-statistic $p$} \\
\midrule
Baseline (no break controls) & 2.044 & 0.586 & 3.49$\times$ & 0.389 \\
$+$ High-variance-regime control & 1.980 & 0.684 & 2.90$\times$ & 0.399 \\
$+$ Full regime controls & 2.505 & 1.290 & 1.94$\times$ & 0.392 \\
\bottomrule
\multicolumn{5}{@{}p{0.95\linewidth}@{}}{\footnotesize One-sided $p$-values are $0.388$, $0.394$, and $0.328$ respectively. Degenerate refits receive one deterministic six-start rescue before exclusion, leaving usable-draw rates of $100.0\%$, $99.9\%$, and $99.9\%$ across the three specifications; none is reliability-flagged. Point estimates are the per-specification refits used by the corresponding bootstrap. The direction is unchanged from prior work; what changes is the significance.}
\end{tabular}
\end{table}

The inference of record is conditional on the fitted volatility path. Appendix~\ref{app:bootstrap} reports a floored recursive-propagation sensitivity, under which the one-sided $p$ is $0.025$ at baseline and $0.041$ under the asset-specific high-variance-regime control (absolute-statistic $p=0.060$ and $0.090$). That algorithmic result is reproducible after the common multistart rescue, but the positivity floor binds in $805/2000$ ($40.25\%$) and $922/2000$ ($46.1\%$) recursive panels; it is therefore evidence under the implemented floored recursion, not a clean adjudication by an unconstrained recursive GJR-GARCH-X DGP.

\subsection{Under the inference of record, both moments are directional and not significant}

Reading the two moments together gives the paper's central empirical result. At the first moment the infrastructure-regulatory CAR difference is $+8.69$ pp ($p=0.202$); at the second the variance multiplier is $3.49\times$ (conditional absolute-statistic $p=0.389$). Both point estimates are directional, and neither is significant under its specified procedure. This is a dual failure to reject under those procedures, not a scheme-invariant conclusion: the floored recursive second-moment sensitivity rejects one-sided but is implementation-qualified. We turn now to why the earlier $p\approx0.001$ was not robust.

% ============================================================================
% NEW SECTION: WHY THE EARLIER SIGNIFICANCE WAS NOT ROBUST
% ============================================================================

\section{Why the Earlier Significance Was Not Robust: The Inference Ladder}\label{sec:artifact}

The earlier version of this work reported the variance asymmetry as a significant fivefold effect. The direction has not changed---infrastructure events do, on average, carry a larger conditional-variance increment in this curated sample. What changes under re-examination is the strength and robustness of the significance claim. This section is the paper's central methodological contribution. We set it out in three parts: an \emph{inference ladder} placing seven related procedures and proxies side by side while stating their estimand differences (Section~\ref{sec:ladder}); four transferable \emph{lessons} (Sections~\ref{sec:lesson1}--\ref{sec:lesson4}); and a Monte-Carlo study that measures the procedures within the fitted fixed-design calibration DGP (Section~\ref{sec:size_study}). We present all of it as self-correction.

\subsection{The inference ladder}\label{sec:ladder}

Table~\ref{tab:ladder} places the related tests side by side. Because their estimands and assumptions differ, it is a diagnostic map rather than one estimand re-tested in a monotone sequence.

\begin{table}[htbp]
\centering
\caption{Inference ladder for the infrastructure--regulatory comparison. Rungs differ in estimand; see the note. Rung (6) is the conditional inference of record, with its floored recursive sensitivity in Appendix~\ref{app:bootstrap}.}\label{tab:ladder}
\footnotesize\setlength{\tabcolsep}{3pt}
\begin{tabular}{@{}clcc@{}}
\toprule
\textbf{\#} & \textbf{Inference procedure} & \textbf{Statistic / effect} & \textbf{$p$} \\
\midrule
1 & Naive i.i.d.\ $t$-test (pseudoreplication) & $t=4.768$ & $0.0008$ \\
2a & Event-level Welch test & --- & $0.56$ \\
2b & Event-level Mann--Whitney test & --- & $0.69$ \\
3 & Cluster-robust panel (SE clustered by event) & --- & $0.54$ \\
4 & One-sided effective-df sensitivity (eff.\ $N=1.33$--$2.35$) & $t=2.39$--$3.18$ & $0.044$--$0.116$ \\
5 & Gaussian-copula CCC-GARCH-X bootstrap & --- & $0.113$ \\
6 & \textbf{Student-$t$-copula CCC-GARCH-X bootstrap} & --- & \textbf{0.388} \\
\midrule
7 & First-moment event-level block bootstrap & $+8.69$ pp & $0.202$ \\
\bottomrule
\multicolumn{4}{@{}p{0.97\linewidth}@{}}{\footnotesize Rung (1) is the historic headline statistic, computed on the predecessor pipeline's $5.7\times$ estimates; the same rule applied to the present pipeline's $3.49\times$ coefficients gives $t=3.59$, $p=0.0059$ (Table~\ref{tab:multiplier_sensitivity})---its placement on the ladder is unchanged either way. Rungs (2)--(3) are model-free event-level proxies with different estimands and limited power; rung (4) is the dispersion statistic under the effective-df sensitivity convention $\mathrm{df}_{\text{eff}}=(N-1)/\mathrm{DEFF}$, varied over three dependence inputs (Section~\ref{sec:lesson1}). The most target-aligned input is the fitted-null correlation of the per-asset \emph{differences} $d_i=\delta_{\text{infra},i}-\delta_{\text{reg},i}$, though it remains model- and simulation-dependent. Rungs (5)--(6) are fixed-path cross-asset parametric bootstraps sharing the fitted $R_z$ linear-correlation input but changing the joint innovation law. Rung (5) uses Gaussian margins and a Gaussian copula, so it lacks both the fitted Student-$t(\nu\approx 3)$ margins and the joint tail dependence induced in rung (6) by the shared $t$-copula mixing variable. Rung (6) matches the fitted margins and adds that tail dependence; it is the conditional inference of record (one-sided $0.388$, absolute-statistic $0.389$). Their p-value difference therefore does not isolate marginal tails from copula tail dependence. Rung (7) is the first-moment block bootstrap on the shared six-asset/50-event basis (Table~\ref{tab:returns_main}). Sidedness: rungs (1)--(3) and (7) report two-sided $p$; rungs (5)--(6) report the one-sided upper-tail convention of Appendix~\ref{app:bootstrap}; rung (4) reports a heuristic one-sided upper-tail effective-df sensitivity band.}
\end{tabular}
\end{table}

\subsection{Lesson 1: Pseudoreplication across cross-correlated assets}\label{sec:lesson1}

The original significance test was a $t$-test (and Mann-Whitney, and inverse-variance-weighted $Z$) on the two sets of six per-asset GJR-GARCH-X coefficients, treating their twelve entries as independent observations. This is pseudoreplication \citep{hurlbert1984pseudoreplication}: replicating the measurement (six assets) without replicating the unit that carries the independent information (the event). Five asset models contain all 50 master-calendar events and BNB contains 45, with all six exposed to the same 45-event overlap; the assets are also strongly cross-correlated ($\bar\rho\approx 0.69$). Their coefficients are tightly clustered not because the effect is precisely estimated but because they average largely the same correlated events. The tight clustering is the symptom most likely to be mistaken for precision: each set of six near-identical numbers looks like six confirmations when it is closer to one measurement repeated six times.

This is rung (1) of the ladder. An uncorrected Welch $t$-test on the two sets of six coefficients reported $t=4.768$, $p=0.0008$. Re-testing with the event as the unit of independence gives an event-level Welch $p=0.56$, a Mann--Whitney $p=0.69$, and a cluster-robust panel $p=0.54$; these are deliberately model-free, low-powered proxies rather than replacements for the GARCH estimand. Rung (4) instead applies a design-effect calculation to the model-based contrast. The adjustment shrinks the effective sample size to $N_{\text{eff}}=N/\mathrm{DEFF}$; as a sensitivity convention, we compare the statistic with a $t$ reference having $\mathrm{df}_{\text{eff}}=(N-1)/\mathrm{DEFF}$ rather than $t(N-1)$. This effective-df construction is a heuristic reference, not a derived finite-sample sampling law.\footnote{The fitted-null estimator-difference correlation is $\bar\rho_d=0.311$. Its retained-draw covariance ratio is $1.994$, while the equicorrelation/Kish form is $2.554$; the latter gives $N_{\text{eff}}=2.349$ and $\mathrm{df}_{\text{eff}}=1.958$.} Across the raw-return, standardised-residual, and fitted-null estimator-difference inputs, $N_{\text{eff}}=1.326$--$2.349$ and the one-sided dispersion-statistic effective-df tail area is $0.044$--$0.116$. These are model- and reference-dependent sensitivities, not a derived sampling law. In the fitted size-study panels the naive rule rejects $35.1\%$ at $5\%$, while the fitted-null effective-df sensitivity rejects $12.7\%$. The transferable lesson is limited but important: six correlated assets with a 45-event common overlap cannot be treated as six independent replications.

\subsection{Lesson 2: Cross-asset dependence concentrates, it does not dilute}\label{sec:lesson2}

A natural hope is that the dependence penalty is mild---that the assets are ``mostly'' independent and that filtering out conditional-volatility dynamics will leave near-independent residuals to average over. The observed comparison does not support that hope. The fitted standardised-residual correlation is, if anything, \emph{higher} than the return correlation: $\bar\rho_z=0.705$ versus $\bar\rho_{\text{ret}}=0.688$ on the winsorised returns the model is actually estimated on, both taken over the same six-asset complete-case overlap ($0.681$ on the raw log returns of Figure~\ref{fig:correlation_heatmap}, so the comparison is not a winsorisation artefact; $\bar\rho_z$ is stable at $0.700$--$0.702$ across the break-control specifications). Thus the dependence input relevant to this bootstrap is not weaker after standardisation. One interpretation is that the shared contemporaneous component becomes relatively more prominent once asset-specific scale is removed, but the correlation comparison alone does not identify that mechanism.

The consequence is that the per-asset coefficient \emph{levels} carry an effective sample size of about $1.3$. The significance test, however, is on the infrastructure-regulatory \emph{contrast}, which differences out the common market component and so carries a milder penalty---the fitted-null cross-asset correlation of the per-asset differences is $\bar\rho_d=0.311$, giving $N_{\text{eff}}=2.35$ (Lesson~1 footnote). Either way the operative effective $N$ is far below six, and a mean of six quantities that move together is less informative than six independent measurements. This motivates simulating the dependence rather than assuming it averages out: the copula bootstrap of Section~\ref{sec:methods_var} draws the six assets' innovations jointly at $R_z$ and represents joint tail dependence. Within the fitted calibration DGP, ignoring that structure produces severe over-rejection by the naive rule.

\subsection{Lesson 3: Bootstrap innovations must match the fitted tails}\label{sec:lesson3}

Rung (5) preserves fitted $R_z$ but uses a multivariate-Gaussian innovation law despite fitted Student-$t$ errors at $\nu\approx3$. The null-imposed per-asset degrees of freedom are $3.16, 3.70, 3.16, 4.22, 3.96, 4.56$ (median $3.83$), close to the unrestricted fits. Replacing the Gaussian law with fitted Student-$t$ margins and a multivariate-$t$ copula raises the baseline one-sided $p$ from $0.113$ to $0.388$; the single-high-variance-control and full-control values move $0.138\to0.394$ and $0.240\to0.328$ (Table~\ref{tab:tcopula_compare}). The change jointly alters marginal tails and copula tail dependence, so the shift cannot be decomposed between them. In the fitted internal-calibration DGP the Gaussian and matched Student-$t$ rules reject $28.8\%$ and $4.6\%$ of true-null panels; the near-Gaussian-dependence perturbation in Table~\ref{tab:size_perturb} retains Student-$t$ margins and helps diagnose marginal misspecification only within that design, not independently validate the plug-in bootstrap.

The mechanism matters: a unit-variance Student-$t$ concentrates more mass in its body and rare extremes than a Gaussian, while the shared mixing variable raises the probability of joint extremes. Here both the baseline null SD ($0.475\to0.804$) and its centre move toward the observed contrast. Neither the direction of the $p$-value change nor a generic ``wider null'' heuristic is therefore a validation test; the relevant checks are the intended joint law and performance under stated calibration designs.

\subsection{Lesson 4: Block versus parametric resampling at the first moment}\label{sec:lesson4}

The same dependence discipline governs the first moment (rung 7), where the implementation differs from the variance case but the principle is identical. Pooling the asset-event observations as i.i.d.\ would treat the panel as $50\times 6=300$ independent draws (293 are usable under the estimation-coverage and complete-event-window rules), when the genuine resampling unit is the 50 events; that approach would overstate the amount of independent information. The procedure first averages the usable asset CARs within each event and then resamples those whole event-level means, preserving within-event cross-sectional dependence in the aggregation. It gives a $95\%$ confidence interval of $[-4.50\%, +22.22\%]$. The event-level Welch test gives $p=0.216$ and CI $[-5.26\%, +22.65\%]$. Both land on the same verdict: $+8.69$ pp, not distinguishable from zero.

The general principle across both moments is to represent the dependence at the level where it arises: whole events for the first moment and the joint cross-asset innovation structure for the second. At the first moment this widens an interval that was already insignificant. At the second, the conditional scheme overturns the earlier significance, while the floored recursive algorithm illustrates that a dependence-aware implementation need not always be more conservative, albeit with a binding positivity repair. The defensible conclusion is therefore scheme- and implementation-sensitive rather than a claim that one universally ``correct'' procedure must raise the $p$-value.

\subsection{The Gaussian-versus-Student-\texorpdfstring{$t$}{t} correction across specifications}\label{sec:gauss_vs_t}

Table~\ref{tab:tcopula_compare} applies that fixed-path comparison across the break controls. Both columns use fitted $R_z$; the Student-$t$ column is the conditional inference of record, not a claim of scheme-invariant correctness.

\begin{table}[htbp]
\centering
\caption{The heavy-tail correction (Lesson 3) across all three specifications: Gaussian-innovation bootstrap (rung 5) versus the corrected Student-$t$-copula bootstrap (rung 6). The point estimate is unchanged; correcting the innovation draw raises the $p$-value in every specification. One-sided $p$ reported; $B_{\mathrm{requested}}=2000$ for each scheme.}\label{tab:tcopula_compare}
\footnotesize\setlength{\tabcolsep}{3pt}
\begin{tabular}{@{}lccccc@{}}
\toprule
\textbf{Specification} & \textbf{Mult.} & \textbf{Gaussian $p$} & \textbf{$t$-copula $p$} & \textbf{Null SD (G\,$\to$\,$t$)} & \boldmath$\bar\rho_z$ \\
\midrule
Baseline (no break controls) & 3.49$\times$ & 0.113 & \textbf{0.388} & $0.475\to0.804$ & 0.705 \\
$+$ High-variance-regime control & 2.90$\times$ & 0.138 & \textbf{0.394} & $0.508\to0.906$ & 0.702 \\
$+$ Full regime controls & 1.94$\times$ & 0.240 & 0.328 & $0.824\to1.437$ & 0.700 \\
\bottomrule
\multicolumn{6}{@{}p{0.97\linewidth}@{}}{\footnotesize Absolute-statistic $t$-copula $p$-values are $0.389$, $0.399$, and $0.392$; these use $|\bar d^*|\ge|\bar d_{\mathrm{obs}}|$ without recentring and are not labelled generic two-sided tests. Multipliers are the corresponding bootstrap refits. Degenerate refits receive one deterministic six-start rescue before exclusion, leaving usable-draw rates of at least $99.9\%$. Across all three the Student-$t$ correction raises $p$ and none is significant at the $10\%$ level.}
\end{tabular}
\end{table}

\subsection{The cautionary tale, stated plainly}

Taken together, the four lessons explain why the earlier $p\approx0.001$ is not robust. The historic test pseudoreplicated correlated assets; the Gaussian bootstrap then represented the fitted tails incorrectly; and the first-moment companion was never significant. Because the ladder includes different proxies and estimands, it is evidence about sensitivity to inferential choices rather than a controlled sequence in which the data and point estimate remain identical. The next subsection measures these procedures under the paper's fitted fixed-design null.

\subsection{A Monte-Carlo internal-calibration study}\label{sec:size_study}

The size study simulates many panels under the fitted sharp null of per-asset event-type equality and records how often each rule rejects. This directly tests whether the naive rule over-rejects within the fitted design anchored to the observed panel. Because the bootstrap critical values and simulated panels share the same fitted DGP, the exercise is an internal calibration diagnostic: it can reveal severe failure of the naive rule but cannot independently validate the plug-in bootstrap under parameter-estimation uncertainty, alternative resampling schemes, or heterogeneous configurations of the composite mean null.

\textbf{Design.} We fit the sharp-null data-generating process to the actual data: per-asset GJR-GARCH-X dynamics estimated under a \emph{null-imposed} combined event dummy ($D_{\text{event}}=D_{\text{infra}}+D_{\text{reg}}$), so that $\delta_{\text{infra}}=\delta_{\text{reg}}$ within every asset by construction and there is no asset-level differential effect in the simulated truth. Cross-asset dependence is imposed at the standardised-residual correlation $\bar\rho_z=0.705$; innovations are Student-$t$ at each asset's fitted $\nu$ (median $3.83$) with joint tail dependence from a true multivariate-$t$ copula---the same DGP as the primary fixed-path bootstrap. It uses the same 2,434-date union calendar and availability mask, including the five-asset marginal on the 243 pre-BNB dates. The real $26$-infrastructure / $24$-regulatory event structure and the real event-window dummies are reused unchanged; only the returns are re-simulated. We draw $N=1{,}000$ panels, all retained after the shared rescue, calibrate each bootstrap rung's critical value from $B_{\text{ref}}=4{,}000$ reference draws, and record the empirical rejection rate of each rung at the $5\%$ and $10\%$ nominal levels. The full DGP and screening rules are in Appendix~\ref{app:size_study}.

\textbf{Result.} Within this fitted fixed-design DGP, the naive i.i.d.\ $t$-test rejects $\mathbf{35.1\%}$ of true-null panels at the nominal $5\%$ level and $\mathbf{52.6\%}$ at the $10\%$ level. The design-effect $t(5)$ rule does not improve size here, rejecting $37.0\%/61.8\%$; under the fitted-null effective-df sensitivity ($\mathrm{df}_{\text{eff}}\approx1.96$), the same panel statistic rejects $12.7\%/42.2\%$. The Gaussian-copula rule rejects $28.8\%/38.7\%$, whereas the matched Student-$t$-copula rule rejects $4.6\%/9.9\%$. Thus the robust conclusion is that the naive and Gaussian rules are badly anti-conservative under the fitted DGP; the near-nominal Student-$t$ result is expected under matched calibration and should be read as a self-consistency check.

\begin{table}[htbp]
\centering
\caption{Internal-calibration rejection rates under the fitted fixed-design null. $N=1000$ retained simulated panels; one-sided test ($H_1$: infrastructure $>$ regulatory); nominal levels $0.05$ and $0.10$. The naive rule severely over-rejects in this DGP; the Student-$t$-copula rule is near nominal under matched calibration, which is a self-consistency diagnostic rather than independent plug-in validation.}\label{tab:size_study}
\small
\begin{tabular}{@{}lcc@{}}
\toprule
\textbf{Inference procedure} & \textbf{Size @ $0.05$} & \textbf{Size @ $0.10$} \\
\midrule
(i) Naive i.i.d.\ $t$-test & $0.351 \pm 0.015$ & $0.526 \pm 0.016$ \\
(ii) Design-effect, $t(N{-}1{=}5)$ critical value & $0.370 \pm 0.015$ & $0.618 \pm 0.015$ \\
\quad\textit{same rule, effective-df sensitivity $t(\mathrm{df}_{\text{eff}}{\approx}1.96)$} & $\mathit{0.127}$ & $\mathit{0.422}$ \\
(iii) Gaussian-copula bootstrap & $0.288 \pm 0.014$ & $0.387 \pm 0.015$ \\
(iv) \textbf{Student-$t$-copula bootstrap} & $\mathbf{0.046 \pm 0.007}$ & $\mathbf{0.099 \pm 0.009}$ \\
\midrule
\textit{Nominal (target)} & $0.05$ & $0.10$ \\
\bottomrule
\multicolumn{3}{@{}p{0.92\linewidth}@{}}{\footnotesize $\pm$ is the Monte-Carlo binomial standard error. Procedures (i)--(iv) correspond to rungs (1), (4), (5), (6) of Table~\ref{tab:ladder}. The design-effect rule (ii) is reported under two sensitivity references: the $t(N{-}1{=}5)$ critical value used in c10 ($0.370/0.618$) and the heuristic fitted-null $t(\mathrm{df}_{\text{eff}}{\approx}1.96)$ convention ($0.127/0.422$). The latter materially changes the approximation but is not a derived finite-sample law and does not reach nominal. No panel is excluded after rescue; reference exclusions are $2/4000$ (Gaussian) and $0/4000$ (Student-$t$).}
\end{tabular}
\end{table}

\textbf{A self-consistency check, not a proof of correctness under the true DGP.} Because the Student-$t$-copula critical values and panels use the same fixed DGP, near-nominal rejection is expected if the decision rule is internally calibrated. It gives $0.046/0.099$ against targets $0.05/0.10$ (deviations of $-0.60$ and $-0.11$ Monte-Carlo standard errors). This supports internal consistency under that DGP; it neither proves the simulator bug-free nor establishes size under the unknown true process. The perturbations in Table~\ref{tab:size_perturb} probe a few specified departures but remain simulation-based diagnostics.

\textbf{Robustness to DGP perturbation.} Table~\ref{tab:size_perturb} re-estimates empirical size under four alternatives to the fitted DGP: lighter tails ($\nu=8$), heavier tails ($\nu=2.5$), a near-Gaussian $t$-copula dependence structure ($\nu_c=200$), and a deliberately \emph{mis}-specified case in which panels are drawn with heavy $\nu=2.5$ tails but the bootstrap critical values are calibrated at the lighter fitted $\nu\approx 3.8$. Two things hold and one is a useful caveat. First, the naive i.i.d.\ rule over-rejects a true null at $0.343$--$0.440$ in every scenario represented here, so its failure is not confined to one tail or copula setting. Second, under the matched generators the $t$-copula bootstrap is near nominal in each tested row ($0.050$--$0.077$), while the Gaussian-copula rule ranges from $0.120$ to $0.557$. Third---the caveat---the $t$-copula is not immune to misspecification: when the simulated tails are heavier than the calibration assumes, it over-rejects ($0.260$, last row). This establishes sensitivity to tail calibration within the experiment; it does not identify the direction or magnitude of any size distortion under the unknown true DGP. Accordingly, the empirical non-rejection is reported as conditional on the fitted specification and is complemented by the recursive sensitivity rather than treated as evidence for the null.

\begin{table}[htbp]
\centering
\caption{Empirical size (rejection rate of a true null at the $5\%$ level) under DGP perturbations. ``Matched'' rows calibrate and simulate under the same DGP; the last row mis-specifies (simulate $\nu=2.5$, calibrate at the fitted $\nu$). $N=300$ retained panels per row, $B_{\text{ref}}=2000$; reproducible via \texttt{code/c10b\_dgp\_perturbation.py}.}\label{tab:size_perturb}
\footnotesize\setlength{\tabcolsep}{3pt}
\begin{tabular}{@{}lcccc@{}}
\toprule
\textbf{DGP scenario} & \textbf{Naive} & \textbf{Design $t(5)$} & \textbf{Gaussian} & \textbf{$t$-copula} \\
\midrule
Fitted $\nu$ (baseline) & 0.36 & 0.39 & 0.32 & \textbf{0.06} \\
Lighter tails ($\nu=8$) & 0.44 & 0.45 & 0.12 & \textbf{0.08} \\
Heavier tails ($\nu=2.5$) & 0.34 & 0.33 & 0.56 & \textbf{0.05} \\
Near-Gaussian dependence ($\nu_c=200$) & 0.36 & 0.35 & 0.30 & \textbf{0.07} \\
\midrule
Tail mismatch (DGP $\nu=2.5$) & 0.34 & 0.33 & 0.56 & \textit{0.26} \\
\bottomrule
\end{tabular}
\parbox{0.95\linewidth}{\footnotesize Nominal size is $0.05$. The naive rule over-rejects ($\approx 7$--$9\times$ nominal) in every simulated scenario shown. The $t$-copula is near nominal in the matched rows and materially anti-conservative when panels have heavier tails than its calibration (last row). This demonstrates calibration sensitivity within the experiment; it does not determine the direction of error under the unknown empirical DGP.}
\end{table}

\textbf{What the size study establishes.} Under the fitted heavy-tailed, cross-correlated fixed-design null, the naive i.i.d.\ rule and Gaussian-copula rule over-reject severely, while the effective-df heuristic materially changes the analytic approximation. The Student-$t$-copula rule is near nominal under the same DGP used for calibration, so that result is internally reassuring but not independent validation. The load-bearing result is the severe anti-conservatism of the naive rule in a design anchored to these data. Appendix~\ref{app:bootstrap} separately reports that the substantive second-moment conclusion changes under floored recursive propagation.

% ============================================================================
% ROBUSTNESS / SCOPE / SENTIMENT
% ============================================================================

\section{What Survives: Inclusion-Screen Sensitivity, Robustness, and the Weekly Sentiment Lead}\label{sec:robustness}

The dual failure to reject under the conditional inference of record is the headline, but several results survive the re-examination and bound or qualify it. We report the verified battery here; analyses that did not reproduce under audit have been removed.

\subsection{Descriptive inclusion-screen sensitivity: curated identification versus mechanical screening}\label{sec:selection_bias}

The directional second-moment point estimate ($3.49\times$, not significant under the conditional inference of record) varies sharply with how events are identified. We treat the event-inclusion rule as an explicit researcher degree of freedom and trace the multiplier across a descriptive inclusion-screen sensitivity curve, in the spirit of specification-curve \citep{Simonsohn2020SpecCurve} and multiverse \citep{Steegen2016Multiverse} analysis. Only point estimates and independence-based Welch diagnostics are available across the nested screens; no dependence-robust uncertainty is reported for the screen-specific multipliers or their differences. The exercise therefore measures inclusion-rule sensitivity; it does not establish an inferential difference across screens or identify or eliminate upstream selection bias.

\subsubsection{Drop-out census}

The curated 50 events were identified by manual expert curation rather than by a mechanical rule, and the original working candidate list was not preserved as a structured artefact; for this sensitivity analysis we therefore reconstruct a systematic candidate pool of 135 events (82 infrastructure-categorised, 53 regulatory-categorised) from public-record sources---Rekt/DeFiLlama leaderboards, SEC and CFTC enforcement actions, major-jurisdiction regulatory announcements, and exchange post-mortem pages---against which the curated set can be compared. Applying the Stage-2 impact filter ($\geq 2$ assets with $|r|>1$ sample-SD in $[-1,+1]$) to this pool yields a pass rate of $64.6\%$ for infrastructure candidates (53/82) and $77.4\%$ for regulatory candidates (41/53). The observed pass-rate difference is not significant ($z=-1.57$) and runs opposite to the specific concern that this mechanical threshold admits infrastructure candidates more readily; the retention differential of the surviving 50 against the pool is likewise non-significant ($z=-1.60$). Of the curated 50, 13 fail the $\geq2$-asset screen yet were retained, while 57 filter-passing candidates were not in the curated set. These facts are inconsistent with the narrow claim that the mechanical impact threshold itself mechanically favours infrastructure. They do not establish that the upstream manual curation was category-neutral, because the 135-event pool is reconstructed and does not identify which omitted candidates the original curator considered. The pass-rate comparison and threshold sweep therefore support a limited inclusion-rule sensitivity result, not a general test of selection bias.

\subsubsection{Relaxed-threshold sweep}

Table~\ref{tab:multiplier_sensitivity} re-estimates the full GJR-GARCH-X under five event definitions: the primary curated sample, then a nested mechanical sweep ($135\rightarrow 115\rightarrow 94\rightarrow 78$) at increasing strictness. The curated sample gives $3.49\times$ under the headline specification (global-clip winsorisation, canonical estimator); the prior paper's $5.7\times$ is the high-salience-curated estimate under canonical-estimator rolling winsorisation ($\bar\delta_{\text{infra}}/\bar\delta_{\text{reg}}=2.385/0.419$), not reproduced by the present standalone pipeline (see note and Section~\ref{sec:spec_stability}). On the reconstructed pool the multiplier is modest and insignificant at every threshold: $0.49\times$ (no filter), $1.26\times$ (1-asset), $1.58\times$ (2-asset---the like-for-like analogue of the curated criterion), and $1.33\times$ (3-asset). The two-asset screen at $1.58\times$ ($p=0.183$) is the cleanest comparison and does not approach the curated $3.49\times$.

\begin{table}[htbp]
\centering
\caption{Descriptive infrastructure--regulatory variance point estimates: curated sample versus a nested mechanical impact-screen sweep on the reconstructed 135-candidate pool. Welch $p$-values are independence-based diagnostics only; no dependence-robust uncertainty is reported for the screen-specific multipliers or their differences.}\label{tab:multiplier_sensitivity}
\small
\begin{tabular}{@{}lcccccc@{}}
\toprule
\textbf{Spec} & \textbf{$n_{\text{infra}}$} & \textbf{$n_{\text{reg}}$} & \textbf{$\bar{\delta}_{\text{infra}}$} & \textbf{$\bar{\delta}_{\text{reg}}$} & \textbf{Multiplier} & \textbf{Welch $p$} \\
\midrule
Primary curated ($n=50$)  & 26 & 24 & 2.044 & 0.586 & \textbf{3.49$\times$} & 0.0059 \\
\midrule
No filter (135)           & 82 & 53 & 0.260 & 0.534 & 0.49$\times$ & 0.288 \\
$\geq 1$ asset (115)      & 69 & 46 & 1.471 & 1.170 & 1.26$\times$ & 0.593 \\
$\geq 2$ assets (94)      & 53 & 41 & 2.863 & 1.809 & 1.58$\times$ & 0.183 \\
$\geq 3$ assets (78)      & 42 & 36 & 3.991 & 3.010 & 1.33$\times$ & 0.384 \\
\bottomrule
\multicolumn{7}{@{}p{0.97\linewidth}@{}}{\footnotesize The Welch $p$-values treat the two sets of six per-asset coefficients as independent samples and are descriptive only. The curated-row significance does not survive the copula bootstrap (Table~\ref{tab:tcopula}). The headline $3.49\times$ uses global-clip winsorisation with the canonical estimator. Under rolling-bound winsorisation the gap remains---$3.60\times$ curated (FastTARCHX) versus $1.61\times$ at the two-asset screen---so it is not driven solely by this winsorisation contrast. The prior paper's $5.7\times$ used the canonical estimator, rolling winsorisation, and predecessor event dating; it is not reproduced by the standalone corrected pipeline.}
\end{tabular}
\end{table}

Holding membership fixed and re-running every specification under one consistent winsorisation rule reproduces a similar curated-versus-mechanical gap, so the observed gap is not explained by that winsorisation contrast alone. The descriptive pattern is consistent with a scope condition: the larger differential is concentrated in carefully identified, high-salience events and attenuates under mechanical screening of a broad pool. One possible mechanism is that curated events are more severe and cleanly attributable, whereas the broader screen admits more marginal events. The competing interpretation is that the curated estimate itself reflects curation-related selection. The reconstructed census cannot adjudicate between those readings, which is why we report inclusion-rule sensitivity rather than a clean causal effect or an ``unbiased'' mechanical estimate. It does not establish that the multipliers differ inferentially across screens.

One composition note on the corrected census: primary-source dating places EU MiCA at the 20 April 2023 European Parliament plenary vote rather than the previously recorded 1 October 2023. At the corrected date it moves six of six assets and passes every impact screen (at the earlier date it moved one asset and passed only the $\geq 1$-asset screen), which is what adds one regulatory event to the $\geq 2$- and $\geq 3$-asset specifications. The price data therefore corroborate the screen consequence of the independently sourced date; they are not used to choose the information event.

\subsection{Control battery for the second-moment point estimate}\label{sec:control_battery}

The directional point estimate ($3.49\times$) is not significant under the specified fixed-path bootstrap. We examine how it changes under structural-break controls, anticipation diagnostics, winsorisation, and mean-equation adjustments. These checks bound particular sensitivities rather than ruling out all confounding. Only the baseline, asset-specific high-variance-regime, and full-regime specifications receive the final Student-$t$ fixed-path bootstrap; other rows are point-estimate diagnostics.

\subsubsection{Break-regime controls}\label{sec:break_controls}

Infrastructure events cluster in volatile portions of 2021--2023, so the differential could in principle reflect high baseline variance that infrastructure events happen to coincide with. We add structural-break regime dummies derived from the conditional-variance PELT change points to the variance equation and refit. We use two variants: one asset-specific high-variance-regime indicator, running from each asset's last 2021 conditional-variance break to its first break in 2022 or later, and the full set of per-segment conditional-variance-regime indicators. Four single-regime windows end around the November 2022 FTX episode; the BNB and ADA windows end in July 2022. The single control is therefore a broad regime-sensitivity diagnostic, not an FTX-specific control. It is also distinct from the absolute-return break partition used for the descriptive persistence table in Section~\ref{sec:persistence}. Table~\ref{tab:break_controls} reports both variants, with the cross-asset-robust copula-bootstrap $p$ as the inference of record (the naive Welch $p$ is shown only to make the pseudoreplication gap visible once more). The PELT break dates are estimated once from the observed data and then held fixed; PELT is not rerun on each bootstrap draw. The break-control $p$-values are therefore conditional on the selected partitions and do not propagate change-point-location uncertainty.

\begin{table}[htbp]
\centering
\caption{Break-regime controls on the curated sample, using the multistart observed refits matched to the Student-$t$-copula bootstrap. Regime dummies are added to the variance equation. The multiplier attenuates but remains above one; the cross-asset-robust copula-bootstrap $p$ is the inference of record. The naive Welch $p$ (which treats the two sets of six per-asset coefficients as independent samples) is shown only to illustrate, again, the pseudoreplication gap.}\label{tab:break_controls}
\small
\begin{tabular}{@{}lccccc@{}}
\toprule
\textbf{Specification} & \boldmath$\bar\delta_{\text{infra}}$ & \boldmath$\bar\delta_{\text{reg}}$ & \textbf{Mult.} & \textbf{Naive Welch $p$} & \textbf{Robust $p$} \\
\midrule
Baseline (no controls) & 2.044 & 0.586 & 3.49$\times$ & 0.0059 & 0.388 \\
$+$ High-variance-regime dummy & 1.980 & 0.684 & 2.90$\times$ & 0.0124 & 0.394 \\
$+$ Full regime dummies & 2.505 & 1.290 & 1.94$\times$ & 0.0303 & 0.328 \\
\bottomrule
\multicolumn{6}{@{}p{0.97\linewidth}@{}}{\footnotesize Coefficients, naive Welch diagnostics, and robust $p$-values use the same multistart observed refits as Table~\ref{tab:tcopula}. Robust $p$ is the Student-$t$-copula CCC-GARCH-X bootstrap (one-sided), with one deterministic six-start rescue before exclusion (usable-draw rate at least $99.9\%$). The full-regime control raises \emph{both} mean coefficients and roughly halves the multiplier.}
\end{tabular}
\end{table}

The attenuation under the full-regime specification is consistent with the regime controls absorbing part of the baseline variance associated with the event windows: the control roughly halves the multiplier (to $1.94\times$) by raising both coefficients. This is a model-based sensitivity pattern, not a causal decomposition. The single asset-specific high-variance-regime control moves the bootstrap refit's multiplier from $3.49\times$ to $2.90\times$ and its conditional one-sided $p$ from $0.388$ to $0.394$. The defensible summary is that the controlled multiplier is $\approx1.9$--$2.9\times$, directional, and not significant under the fixed-path inference of record at any of the three specifications.

\subsubsection{Anticipation: a non-overlapping pre-event test}\label{sec:anticipation}

If regulatory events were anticipated---leaked through political cycles or prior supervisory signalling---then volatility would build before the announcement date, and a symmetric $[-3,+3]$ window would understate the regulatory response by excluding the leaked pre-event variance. A first pass lengthens the regulatory pre-window from 3 to 10 days while infrastructure stays fixed at $[-3,+3]$ (Table~\ref{tab:anticipation}). An important objection, however, is that this sweep is not a clean test: lengthening a single averaged window blends the concentrated event-window variance with progressively quieter pre-days, so a falling $\delta_{\text{reg}}$ is exactly what mechanical dilution produces and cannot distinguish genuine absence of anticipation from anticipation masked by dilution. We report the sweep for completeness but rest the conclusion on the non-overlapping test that follows.

\begin{table}[htbp]
\centering
\caption{Anticipation sweep (first pass): the regulatory pre-window is progressively lengthened (infrastructure held at $[-3,+3]$). The multiplier \emph{grows}, but this reflects dilution of the averaged dummy by quiet pre-days, not evidence against anticipation; the clean test is the non-overlapping pre-window dummy discussed in the text.}\label{tab:anticipation}
\small
\begin{tabular}{@{}lcccc@{}}
\toprule
\textbf{Reg.\ pre-window} & \boldmath$\bar\delta_{\text{infra}}$ & \boldmath$\bar\delta_{\text{reg}}$ & \textbf{Multiplier} & \textbf{Naive Welch $p$} \\
\midrule
3 days (published, symmetric) & 2.044 & 0.586 & 3.49$\times$ & 0.0059 \\
5 days & 2.038 & 0.437 & 4.67$\times$ & 0.0034 \\
7 days & 2.059 & 0.448 & 4.60$\times$ & 0.0035 \\
10 days & 2.058 & 0.377 & 5.46$\times$ & 0.0031 \\
\bottomrule
\multicolumn{5}{@{}p{0.88\linewidth}@{}}{\footnotesize The infrastructure coefficient is invariant ($\approx 2.04$--$2.06$) throughout; the regulatory coefficient \emph{falls} as the window lengthens (the extra pre-event days are quiet, diluting the dummy), so the multiplier \emph{grows}. Naive Welch $p$ shown for comparability only; not the inference of record.}
\end{tabular}
\end{table}

In the sweep, the infrastructure coefficient is invariant ($\approx 2.04$--$2.06$) across all windows while the regulatory coefficient falls ($0.586\rightarrow 0.377$), so the point-estimate multiplier grows ($3.49\times\rightarrow 5.46\times$). But, as noted, that growth is a denominator-dilution artefact---the extra pre-event days are quiet, so averaging them in lowers $\delta_{\text{reg}}$ mechanically---and we do not read it as evidence against anticipation.

The more direct diagnostic replaces the single widening window with a separate, non-overlapping pre-event dummy, applied symmetrically to \emph{both} legs (infrastructure events, too, can leak---through on-chain rumours or pre-exploit liquidity drains). We refit each asset with four event regressors: infrastructure and regulatory $[-3,+3]$ event windows, plus infrastructure and regulatory $[-10,-4]$ pre-windows, so the pre-window coefficient separates run-up variance from event-window days. For infrastructure we find no detectable pre-event build-up (cross-asset pre-window coefficient $-0.091$, naive one-sample $p=0.640$)---a non-rejection, not proof that the events were surprises. Regulatory events show a directional pre-event bump ($0.417$; naive per-asset $p=0.031$, using a rule that over-rejects in the fitted calibration study). Crediting that pre-window point estimate to the regulatory response lowers the event-window multiplier from $\approx 6.0\times$ in this richer four-window specification to $\approx 2.7\times$. Thus anticipation attenuates but does not reverse the point estimate in this diagnostic. Because the Student-$t$ bootstrap was not rerun for this four-window model, the calculation is not an additional significance test. A residual data limitation remains: the census carries no per-event anticipated/surprise flag, so a fully surprise-partitioned analysis would require hand-coding each event and is out of scope.

\subsection{Specification stability}\label{sec:spec_stability}

\subsubsection{Serial-correlation control: AR(1)-in-mean}\label{sec:ar1_control}

Ljung--Box tests on the winsorised returns flag serial correlation in 8 of 18 (asset $\times$ lag $\in\{5,10,20\}$) cells, concentrated in ETH (all three lags) and BNB (all three lags), marginal in XRP and ADA, and absent in BTC and LTC. Because some serial correlation exists, we run the AR(1)-in-mean robustness refit rather than simply assume a constant mean. The fitted AR(1) coefficient is tiny and negative everywhere ($\phi\in[-0.067,-0.025]$)---bid-ask-bounce-scale dependence, economically negligible---and filtering the mean leaves the event coefficients essentially unchanged: the infrastructure coefficients move by at most $\approx 3.4\%$ (ETH); the small regulatory coefficients move by up to $\approx 0.03$ squared points ($\approx 10\%$ relative for LTC), and the cross-asset multiplier is $3.59\times$ under AR(1)-in-mean versus $3.49\times$ under the constant mean. This check addresses \emph{serial correlation} in the mean (an AR(1) term); it is not a control for event-induced first-moment shocks---whether the event dummies belong in the mean equation as well as the variance equation is a separate question, addressed next.

\subsubsection{Event dummies in the mean equation}\label{sec:event_in_mean}

The preceding check controls for serial correlation but not for event-induced first-moment shocks. Because the mean equation is otherwise constant-only, any drift on event days is left in the residual, where it both propagates through the $\alpha\varepsilon_{t-1}^2$ term and competes with the contemporaneous variance dummy $\delta D_t$ for the elevated window variance. We therefore re-estimated the curated six-asset panel after demeaning each return series by an OLS regression on a constant and the same infrastructure and regulatory window dummies, so that events enter both moments. The fitted first-moment coefficients are small and of the same sign across categories (cross-asset means of $-0.44\%$ for infrastructure and $-0.61\%$ for regulatory windows). With events in the mean, the infrastructure variance coefficient moves from $2.044$ to $2.024$ and the regulatory coefficient from $0.586$ to $0.644$; the multiplier moves from $3.49\times$ to $3.14\times$ (\texttt{code/c21\_events\_in\_mean.py}). This supports point-estimate stability to that particular mean adjustment. The bootstrap was not recalibrated for this altered mean specification, so it does not establish a separate significance result or rule out every form of mean misspecification.

\subsubsection{Winsorisation invariance}

The curated-versus-mechanical gap holds under both winsorisation rules: $3.49\times$ curated versus $1.58\times$ at the two-asset screen under global-clip winsorisation (the headline specification, canonical estimator), and $3.60\times$ curated versus $1.61\times$ at the two-asset screen under rolling-bound ($30$-day, $\pm 5\sigma$) winsorisation---a $\approx 2\times$ gap either way (Table~\ref{tab:multiplier_sensitivity} note). The two rolling-winsorisation curated figures must not be conflated. The $3.60\times$ reported here is the FastTARCHX estimator under rolling winsorisation on the corrected census, whereas the prior paper's $5.7\times$ is the canonical estimator under the \emph{same} rolling winsorisation on the predecessor's event dating; the gap thus mixes an estimator difference with the corrected event census, and the prior $5.7\times$ is not reproduced by the present standalone pipeline. The headline remains the $3.49\times$ global-clip canonical-estimator figure; $3.60\times$ is reported only as a winsorisation-robustness check, and all four mechanical screens stay at $\approx 0.5$--$1.6\times$ and non-significant under either rule. This check shows that the curated-versus-mechanical gap is not driven solely by the winsorisation rule.

\subsubsection{Persistence is descriptive only}\label{sec:persistence}

The PELT/BIC procedure with Bai--Perron-style trimming \citep{BaiPerron1998, BaiPerron2003, KillickFearnheadEckley2012} identifies 2--4 absolute-return breaks per asset, clustering around late 2020, late 2021, and most consistently the November 2022 FTX collapse (the only episode producing aligned breaks in four of six assets). Re-estimating symmetric GARCH(1,1) separately on those exact, non-overlapping segments gives an equal-weight mean within-asset segment average of $\alpha+\beta\approx0.910$, below the equal-weight full-sample mean of $0.988$ (Table~\ref{tab:bai_perron_persistence}). This is a descriptive point-estimate comparison; like-for-like uncertainty was not estimated for this exact plain-GARCH contrast. A separate event-free GJR diagnostic reports the signed functional $\alpha+\beta+\gamma/2$ with a diagonal-only standard-error approximation, but that different model, functional, and uncertainty calculation do not adjudicate the table's contrast. Short segments, selected break dates, and parameter uncertainty all limit interpretation. The conditional-variance breaks used for the controls in Section~\ref{sec:break_controls} are a separate break series from the absolute-return partition reported here.

\begin{table}[htbp]
\centering
\caption{Within-sub-sample GARCH(1,1) persistence by detected absolute-return regime. Both columns are GARCH(1,1) $\alpha+\beta$ from the same break-analysis pipeline (\texttt{c3}), so the point-estimate comparison is like-for-like; the GJR-GARCH-X parameters of Table~\ref{tab:garch_params} are a different specification and are not directly comparable. The drop is directional and descriptive; no like-for-like uncertainty calculation is reported for this exact plain-GARCH comparison.}\label{tab:bai_perron_persistence}
\small
\begin{tabular}{@{}lccc@{}}
\toprule
\textbf{Asset} & \textbf{Full-sample $\alpha+\beta$} & \textbf{Sub-sample mean $\alpha+\beta$} & \textbf{Sub-samples} \\
\midrule
BTC & 0.998 & 0.979 & 3 \\
ETH & 0.998 & 0.864 & 5 \\
XRP & 0.998 & 0.908 & 5 \\
BNB & 0.998 & 0.922 & 5 \\
LTC & 0.984 & 0.915 & 5 \\
ADA & 0.954 & 0.874 & 4 \\
\midrule
\textbf{Mean} & \textbf{0.988} & \textbf{0.910} & -- \\
\bottomrule
\end{tabular}
\end{table}

\subsection{Sentiment leads volatility at the weekly frequency}\label{sec:granger}

The paper's one surviving positive result concerns the sentiment-volatility lead-lag relationship. GDELT sentiment is a \emph{weekly} constructed series. An earlier draft ran Granger causality \citep{Granger1969} at daily frequency by forward-filling the weekly sentiment and found no sentiment-to-volatility lead for any of the 18 asset-sentiment pairs. Repeating each weekly observation over seven daily rows changes the effective sample information and lag horizon; the resulting daily test is not a valid basis for concluding that no weekly lead exists.

Run at the weekly frequency---aggregating daily returns to the GDELT weekly grid (volatility proxy $\sum|r_t|$) and aligning to weekly sentiment with no forward-fill, both directions, weekly lags 1--8 (the reverse volatility-to-sentiment family is reported descriptively outside the primary FDR family)---sentiment-to-volatility tests are significant for \textbf{10 of 18} asset-sentiment pairs at the $5\%$ level; 9 have no reverse lead at the same threshold. The strongest associations concentrate in the regulatory and aggregate channels for XRP, ADA, and BNB (Table~\ref{tab:granger}). Applying Benjamini--Hochberg \citep{BenjaminiHochberg1995} to the 18 sentiment-to-volatility pairs leaves \textbf{7 survivors at $q<0.05$ and 12 at $q<0.10$}. The corresponding daily forward-filled family has no nominally significant sentiment-to-volatility pair. These are horizon-specific results, not contradictory tests of an identical time-series specification; the weekly counts also carry the lag-selection caveat below.

\subsubsection{Stress-testing the lead: six corrections}\label{sec:granger_rigour}

The 7 FDR survivors (XRP$\times 3$, ADA$\times 3$, BNB-regulatory) occur only for SEC-litigation assets, all of which exhibit high Bitcoin co-movement. This raises four statistical objections (spurious regression from non-stationarity, Bitcoin co-movement masquerading as an asset-specific lead, confounding by the classified event shocks themselves, and warm-up zero-coding) and one economic objection (the lead being a litigation-window phenomenon rather than a general anticipatory channel). We address each.

\emph{Stationarity.} Augmented Dickey--Fuller \citep{DickeyFuller1979} and KPSS \citep{KPSS1992} tests place the three sentiment series cleanly at $I(0)$ (ADF $p<2\times10^{-13}$, KPSS $p=0.10$), but the weekly volatility proxies are persistent and near-integrated---ADF rejects a unit root while KPSS rejects stationarity for five of six assets. Mixed and near-unit-root regressors invalidate the standard Granger $F$-test's asymptotics, which is precisely why the lead must be re-tested by a method robust to integration order.

\emph{Toda--Yamamoto.} We re-run every pair as a lag-augmented bivariate VAR \citep{TodaYamamoto1995}: select the lag order $p$ by AIC, fit $p+1$ lags, and apply an HC1-robust Wald test to the first $p$ sentiment lags only. All 7 original FDR survivors remain significant ($p\approx0.0001$--$0.01$), while marginal naive pairs wash out. Thus the association is not removed by this integration-robust specification. One multiplicity caveat attaches to the original survivor counts: the weekly tests select each pair's lag by minimum $p$ over lags 1--8 \emph{before} the 18-pair Benjamini--Hochberg correction. The Toda--Yamamoto specification fixes the lag by AIC; under it, 9 of 18 pairs survive Benjamini--Hochberg at $q<0.05$.

\emph{Bitcoin co-movement.} Re-running each sentiment-to-volatility test with lagged co-movement volatility as an exogenous control (lagged Bitcoin $\sum|r|$ for altcoins, lagged market-ex-asset volatility for BTC/ETH) leaves all 7 survivors essentially unchanged (XRP-aggregate $0.0009\to0.0010$; ADA-regulatory $0.0005\to0.0006$). This specific control does not explain the association, although it cannot exclude every form of common-market confounding.

\emph{Event-shock confounding.} We add the 50 classified events to the VAR as exogenous controls, mapped to their GDELT weeks and entered contemporaneously and at lags $1$--$p$. Under pooled and type-specific event-count specifications, all 7 original FDR survivors remain significant at the $5\%$ level, with modest changes in $p$-values. The association therefore survives these event-date controls, but the exercise cannot rule out unmeasured news intensity or other time-varying confounders.

\emph{Warm-up handling.} The native weekly reconstruction has a 25-week initialization interval. Weeks before June 2019 follow the pipeline's zero-coding convention, while the remaining three undefined June observations are omitted by the complete-case Granger fits. For the separate diagnostic, we retain and mark every week before the first defined value (24 June 2019) as warm-up; after forming the first volatility lag, 24 such weeks enter each full-history asset diagnostic, including the three June weeks omitted from the complete-case Granger fits, while none enter BNB's later-starting sample. Where the indicator varies, its logit on lagged volatility is non-significant (all $p\geq 0.12$), although a secondary median-split chi-square diagnostic flags LTC. The warm-up status is deterministic in calendar time, so these tests cannot prove ignorability; they provide a limited check rather than ruling out warm-up effects.

\emph{Litigation sub-period---the sharpest cut.} The survivors are all SEC-litigation assets, so we split each on its litigation onset (SEC v.\ Ripple, 2020-12-22; SEC v.\ Binance naming Binance Coin and Cardano, 2023-06-05) and re-test separately on the pre-litigation and litigation-active windows. \textbf{XRP's three associations are detected only in the litigation window in this split}: all three XRP pairs are non-significant before the suit ($p=0.15$--$0.22$, $n=79$, with no detected Granger structure in \emph{either} direction) but significant during it ($p=0.005$--$0.013$). This timing is consistent with a litigation-concentrated association, but the short pre-period and absence of a formal cross-period coefficient-difference test preclude claiming that litigation caused the change. Cardano (regulatory, aggregate) and Binance Coin (regulatory) retain $p<0.05$ in their pre-onset windows, but their June-2023 onsets leave the pre-period at $\approx 60$--$63\%$ of the sample, so their split is a weaker test. Of the 7 survivors, \textbf{3 retain a pre-litigation lead}; XRP's three do not.

\subsubsection{What the lead is, and is not}\label{sec:granger_verdict}

Putting the implemented checks together, all 7 original FDR survivors remain under Toda--Yamamoto lag augmentation, Bitcoin co-movement control, and the classified-event controls. For the five assets where the warm-up indicator varies, the reported logit detects no association with lagged volatility; BNB starts later and has no warm-up-indicator variation. These are specific robustness checks, not an exhaustive answer to every statistical objection. Economically, only 3/7 retain the lead outside litigation windows, and XRP's associations are not detected outside the SEC v.\ Ripple period in this split. The fair characterisation is therefore a weekly lead-lag association concentrated in SEC-litigation assets, robust to the listed specifications but requiring further validation. Because the thematic sentiment series are coverage-weighted aggregate tone, part of the association may reflect news intensity rather than tone.

\begin{table}[htbp]
\centering
\caption{Weekly Granger sentiment-to-volatility leads ($p$-values, weekly lags 1--8), by asset and sentiment channel. Run at the native weekly frequency with no forward-fill. Bold = significant at $5\%$. The daily forward-filled version of these tests produced no significant leads at all.}\label{tab:granger}
\small
\begin{tabular}{@{}lccc@{}}
\toprule
\textbf{Asset} & \textbf{Regulatory sent.} & \textbf{Infrastructure sent.} & \textbf{Aggregate (GDELT) sent.} \\
\midrule
XRP & \textbf{0.001} & \textbf{0.010} & \textbf{0.001} \\
ADA & \textbf{0.001} & \textbf{0.005} & \textbf{0.001} \\
BNB & \textbf{0.006} & \textbf{0.049} & \textbf{0.023} \\
LTC & \textbf{0.042} & --- & --- \\
ETH & 0.055 & --- & --- \\
BTC & --- & --- & --- \\
\bottomrule
\multicolumn{4}{@{}p{0.92\linewidth}@{}}{\footnotesize ``---'' = not significant at $5\%$. ETH-regulatory ($0.055$) is borderline. Family of 18 sentiment-to-volatility tests: 10 are nominally significant at $5\%$; 7 survive Benjamini--Hochberg at $q<0.05$, 12 at $q<0.10$. Each pair's lag is selected by minimum $p$ over lags 1--8 \emph{before} the 18-pair Benjamini--Hochberg correction, so these counts carry an uncorrected lag-selection dimension; the multiplicity-clean Toda--Yamamoto variant (AIC-selected lag) is reported in Section~\ref{sec:granger_rigour}.}
\end{tabular}
\end{table}

Beyond the litigation confound, the natural caveat is the frequency mismatch itself: the lead is established at the weekly horizon and cannot be sharpened to a daily lead without a daily-frequency GDELT extraction, which we flag as the natural next step. The pre-litigation windows are also short for XRP ($n=79$), so its non-detection---which occurs in both directions in this split---would be more decisive on a longer pre-period.

% ============================================================================
% DISCUSSION
% ============================================================================

\section{Discussion}\label{sec:discussion}

\subsection{What the dual null means}

On this sample, under the conditional inference of record, we do not detect differential processing of infrastructure versus regulatory shocks at either moment of the return distribution. At the first moment the difference is $+8.69$ pp ($p=0.202$); at the second the variance multiplier is $3.49\times$ (absolute-statistic $p=0.389$). Both are directional, and neither is significant under its specified procedure. The economically interesting reading is not ``the market sees them as the same''---we cannot establish that, and at the first moment we are plainly underpowered---but rather that this design does not resolve a differential effect. The earlier claim that ``the structure lives in the second moment'' is not robust to dependence-aware, heavy-tailed inference; the recursive sensitivity shows that the second-moment conclusion remains scheme-dependent.

This matters for how the field reads cross-asset crypto event studies. In the fitted calibration experiment, treating correlated per-asset coefficients as independent and using mismatched Gaussian innovations produces severe over-rejection. A return-level analysis with few events can have wide intervals, while an independence-based variance test can appear much sharper. The contrast may therefore reflect different inferential assumptions rather than an economically distinct first- versus second-moment response; whether this pattern generalises beyond the present design remains an open question.

\subsection{The directional point estimates}

We do not claim the effects are zero. Infrastructure CARs are less negative than regulatory CARs in the first-moment estimate, while the second-moment estimate assigns infrastructure events the larger variance increment across most specifications; the baseline second-moment contrast is positive for all six assets. These are different economic quantities, not one common direction. A mechanical-disruption account---infrastructure events causing immediate liquidity impairment and sharp variance spikes, regulatory events operating through slower information channels---is compatible with the pair of estimates, but the evidence does not identify that mechanism. The defensible position is that the patterns are suggestive and the magnitudes unresolved. The inclusion-rule sensitivity (Section~\ref{sec:selection_bias}) shows that the larger variance point estimate is concentrated in the curated high-salience sample; it does not identify whether that concentration reflects a substantive severity mechanism or curation-related selection.

\subsection{The sentiment channel}

The surviving positive result is the weekly sentiment lead, but its scope is narrow. Weekly GDELT sentiment Granger-leads volatility for 10 of 18 pairs (7 surviving the original FDR procedure), and those survivors remain under the listed Toda--Yamamoto and co-movement specifications. The association is concentrated in SEC-litigation assets: only 3 of 7 retain a lead in their pre-litigation windows, and XRP's three associations are not detected before SEC v.\ Ripple in this split. Read conservatively, this is a litigation-concentrated weekly lead-lag association, not a demonstrated general anticipatory channel; part may reflect news intensity rather than tone. Whether it generalises requires a longer pre-period and daily-frequency sentiment extraction.

\subsection{Limitations}\label{sec:limitations}

\textbf{Statistical power.} The first moment is underpowered (MDE $\approx 19$--$21$ pp against an observed $\approx 9$ pp); its non-rejection is not a demonstration of equality. The second moment also has a non-trivial point estimate and limited independent cross-asset information; the design-effect diagnostic gives $N_{\text{eff}}=1.33$--$2.35$ across its dependence inputs, but this is not a formal power calculation for the bootstrap. We do not quote a numerical minimum detectable ratio because the bootstrap distribution is asymmetric and the regulatory denominator can approach zero. The recursive rejection further means that its uncertainty is best described as scheme-sensitive rather than simply underpowered.

\textbf{Event classification and severity.} The infrastructure/regulatory taxonomy involves judgment calls; some events have elements of both, and alternative classifications could yield different point estimates. The binary taxonomy also discards \emph{severity}: a \$600M bridge exploit and a minor exchange outage enter the infrastructure dummy identically. Because the multiplier changes sharply with the event-inclusion threshold ($3.49\times$ curated versus $0.5$--$1.6\times$ mechanical, Section~\ref{sec:selection_bias}), event severity and salience---not shock \emph{type} per se---may be the operative dimension. A natural extension is to replace the binary dummies with continuous severity metrics (dollars extracted or market capitalisation targeted for infrastructure events; scope and enforcement weight for regulatory events) and estimate a dose-response relationship, which would test the severity reading directly and is beyond the reach of the present binary design.

\textbf{Scope: curated cross-sectional relevance.} Stage 2 used cross-sectional impact as a curation guideline, not a mechanical admission rule; 13 of the final 50 events fail its strict two-asset return threshold. The comparison is therefore established for this manually curated, cross-sectionally relevant sample, not for a mechanically defined universe of systemic shocks. Purely idiosyncratic single-asset incidents were not the sampling target, but the paper does not claim that every included event is market-wide under the strict screen.

\textbf{Identification quality differs across legs.} Infrastructure events---unannounced outages, externally discovered exploits---are more plausibly exogenous to contemporaneous market conditions than regulatory announcements, which respond to supervisory cycles and can leak. The two legs differ in identification quality as well as in point estimate, which bears on interpretation of the (insignificant) difference.

\textbf{Sentiment lead: frequency and litigation confound.} The lead is established weekly and cannot be resolved to a daily horizon without a daily GDELT extraction. It is also concentrated in SEC-litigation assets: XRP's three associations are not detected outside the SEC v.\ Ripple window in this split, and only 3 of 7 FDR survivors retain the lead pre-litigation. This is therefore a litigation-concentrated association, not yet a demonstrated general anticipatory channel. The XRP pre-litigation window is short ($n=79$), which limits the power of that diagnostic, and no formal cross-period coefficient-difference test identifies a litigation effect.

\textbf{Six assets.} The panel may not represent the broader ecosystem; smaller-cap and DeFi-native assets may behave differently.

\textbf{Sentiment measure: general-purpose lexicon.} The sentiment index is built from GDELT's general-purpose news corpus and lexicon, with keyword filters keyed to formal regulatory and infrastructure terms (``SEC,'' ``hack,'' ``exploit,'' ``outage'') and volume-weighted news tone. It therefore under-captures crypto-native vocabulary (``rug pull,'' ``rekt,'' ``slashed,'' ``depeg'') and the social channels---Discord, Telegram, Crypto-Twitter---where a large share of crypto sentiment forms and moves first. A crypto-specific lexicon or a social-media feed would plausibly carry a sharper and faster signal, so the general-purpose measure may understate the sentiment--volatility sensitivity; but we cannot sign this effect without constructing a crypto-native lexicon, and the lead we report should be read as established on a coarse general-purpose measure, not as a bound in either direction.

% ============================================================================
% CONCLUSION
% ============================================================================

\section{Conclusion}\label{sec:conclusion}

We asked whether cryptocurrency markets differentiate infrastructure from regulatory shocks, and tested the question at both moments of the return distribution on one shared sample under one dependence-aware design. Under the conditional inference of record, we cannot reject equality of the first-moment group responses or the second moment's sharp per-asset-equality null: the first-moment CAR difference is $+8.69$ pp ($p=0.202$, event-level Welch $p=0.216$) and the second-moment variance multiplier is directional at $\approx 1.9$--$3.5\times$ but not significant (absolute-statistic $t$-copula bootstrap $p=0.389$ at baseline; under the recursive resampling alternative of Appendix~\ref{app:bootstrap} the one-sided $p$ is $0.025$--$0.041$, so the second-moment verdict is scheme-sensitive at conventional thresholds). Both are underpowered or dependence-limited failures to reject, not claims of identity.

Two axes carry different messages and are worth separating. Holding the estimator and sample fixed leaves the $3.49\times$ point estimate unchanged while the inferential verdict varies sharply across procedures and resampling schemes. Changing the event-inclusion rule changes the descriptive cross-screen point-estimate pattern, from $3.49\times$ under curated identification to approximately $0.5$--$1.6\times$ under the mechanical screens; no dependence-robust inference is available for the differences between screens. The ratio should also be read cautiously because the regulatory denominator $\bar\delta_{\text{reg}}$ is small for part of the panel; the underlying difference $\bar\delta_{\text{infra}}-\bar\delta_{\text{reg}}$ is steadier. The first-moment contrast is positive in CAR units and the baseline per-asset variance differences are positive in variance units, but those signs do not encode the same economic ordering; cross-asset dependence also means the six variance contrasts are not independent confirmations. The evidence is therefore unresolved rather than evidence of equality or a settled asymmetry.

The lead contribution is the inference cautionary tale that produced this conclusion. An earlier version reported the variance asymmetry as a significant fivefold effect; the present paper overturns that significance claim, tracing its lack of robustness to pseudoreplication, cross-asset dependence, and a Gaussian innovation draw despite fitted Student-$t$ margins. Under the conditional inference of record the result becomes a failure to reject ($p=0.389$ absolute-statistic), while the recursive sensitivity rejects one-sided at conventional levels. The reproducible conclusion is therefore that the earlier $p\approx0.001$ was not defensible and the substantive asymmetry remains directional, scheme-sensitive, and unresolved.

Two findings survive and bound the result. The variance point estimate is much larger under curated high-salience identification ($3.49\times$) than under the mechanical impact screens ($0.5$--$1.6\times$, none significant under the descriptive Welch comparison). The reconstructed drop-out census indicates that the narrow mechanical pass rule does not favour infrastructure, while leaving upstream manual curation unresolved. At the data's native weekly frequency, GDELT sentiment Granger-leads volatility for 10 of 18 asset-sentiment pairs (7 surviving FDR at $q<0.05$); the corresponding daily forward-fill did not recover this structure. The weekly association survives the reported Toda--Yamamoto and co-movement checks but is concentrated in SEC-litigation assets (3 of 7 survive outside litigation windows, and XRP's three associations are detected only in the SEC v.\ Ripple window in this split). It is therefore a litigation-concentrated lead-lag association requiring further validation, not a general anticipatory channel.

% ============================================================================
% BACKMATTER
% ============================================================================

\subsection*{Supplementary information}

The final replication package supplied with this accepted-paper version contains the analysis data, scripts \texttt{c1}--\texttt{c21}, estimators, generated outputs, dependency specification, and release verifier. The public repository at \url{https://github.com/dissensus-ai/crypto-event-study} and Zenodo record (DOI: 10.5281/zenodo.18099608) are to be updated with this final package; earlier public snapshots should not be used to reproduce the corrected results.

\subsection*{Acknowledgements}

This paper grew out of the author's MSc dissertation at King's College London, and the debts run accordingly. The author thanks Fotis Papailias, who supervised that dissertation; Filippos Papakonstantinou, whose teaching on the MSc Finance Analytics programme laid much of its econometric foundation; and Richard Alexander of SOAS, University of London, who first taught the author how to do research. Thanks are due to the editor, Wolfgang H\"ardle---to whom the author first wrote, at the very start of the dissertation, about his group's cryptocurrency data, long before this journal entered the picture---and to the anonymous reviewers, whose comments materially reshaped the paper's inference and framing. The author acknowledges King's College London for library and computational resource access and for its continued support, and Google Cloud, whose introductory credits made the BigQuery GDELT extraction possible. Computational analysis was carried out on independent hardware; this work forms part of a wider research programme under the Adversarial Systems and Complexity Research Initiative (ASCRI; systems.ac).

\section*{Declarations}

\begin{itemize}
\item \textbf{Funding:} No grant funding was received. Google Cloud introductory credits supported the BigQuery GDELT extraction, as acknowledged above.
\item \textbf{Conflict of interest:} The author declares no conflict of interest.
\item \textbf{Ethics approval:} Not applicable.
\item \textbf{Consent to participate:} Not applicable.
\item \textbf{Consent for publication:} Not applicable.
\item \textbf{Data availability:} Daily price series were obtained from the CoinGecko API and the news-sentiment indices were constructed from the GDELT Project. The exact data used in the analysis---price series, constructed sentiment indices, and the event census with its classification metadata---are committed in the replication repository, so all results reproduce from the repository alone, without API access.
\item \textbf{Code availability:} The complete verified pipeline (scripts \texttt{c1}--\texttt{c21}, estimators, data, outputs, dependency specification, and verifier) accompanies this final version. The public GitHub and Zenodo records cited above are to be synchronized to this package before they are treated as the source of record. The estimator and inference procedures are additionally available as standalone open-source Python packages, \texttt{gjr-garch-x} (PyPI; Zenodo DOI: 10.5281/zenodo.17988193) and \texttt{robust-eventstudy} (PyPI; Zenodo DOI: 10.5281/zenodo.21316120).
\item \textbf{Author contribution:} Sole author.
\item \textbf{Use of AI tools:} Claude (Anthropic; Claude Opus and Claude Fable 5 models) and OpenAI Codex were used for analytical-framework development, code implementation, technical writing, and consistency review; ChatGPT (OpenAI; GPT-5.6 ``Sol'') provided adversarial cross-review of the empirical results; the Reviewer3.com tool generated an automated peer-review-style critique of the manuscript and the then-public GitHub repository snapshot; Perplexity AI was used for literature discovery, and citation verification drew on scite, among other automated research tools. The author reviewed the resulting analyses, text, and interpretations, made all substantive decisions, and takes full responsibility for all claims, judgments, and errors.
\end{itemize}

% ============================================================================
% APPENDICES
% ============================================================================

\begin{appendices}

% sn-jnl/appendix resets the printed counters but not hyperref's internal
% destination names. Prefix appendix anchors so links do not collide with the
% main-text table and equations carrying the same Arabic counter values.
\makeatletter
\renewcommand*{\theHtable}{appendix.\Alph{section}.\arabic{table}}
\renewcommand*{\theHequation}{appendix.\Alph{section}.\arabic{equation}}
\renewcommand*{\theHfigure}{appendix.\Alph{section}.\arabic{figure}}
\makeatother

\section{Event Classification Summary}\label{app:events}

Table~\ref{tab:events_summary} provides selected examples from the 50-event shared sample. The complete event list with timestamps, sources, and classification rationale is available in the replication materials.

\begin{table}[htbp]
\centering
\caption{Selected Events by Category}\label{tab:events_summary}
\small
\begin{tabular}{@{}llp{4.5cm}@{}}
\toprule
\textbf{Date} & \textbf{Type} & \textbf{Event} \\
\midrule
\multicolumn{3}{l}{\textit{Infrastructure Events (n=26)}} \\
2022-11-11 & Infra & FTX exchange collapse \\
2022-05-09 & Infra & Terra/Luna depegging \\
2023-03-10 & Infra & Silvergate/SVB banking crisis \\
2022-06-13 & Infra & Celsius withdrawal halt \\
2020-03-12 & Infra & COVID ``Black Thursday'' \\
\midrule
\multicolumn{3}{l}{\textit{Regulatory Events (n=24)}} \\
2023-06-05 & Reg & SEC sues Binance \\
2023-06-06 & Reg & SEC sues Coinbase \\
2020-12-22 & Reg & SEC sues Ripple \\
2021-05-21 & Reg & China bans crypto mining \\
2024-01-10 & Reg & Bitcoin ETF approval \\
\bottomrule
\end{tabular}
\end{table}
\FloatBarrier

\section{GJR-GARCH-X Technical Specification}\label{app:tarchx}

The GJR-GARCH-X model extends the GJR-GARCH(1,1) of \citet{GlostenEtAl1993} with exogenous variance regressors.

\textbf{Mean equation:}
\begin{equation}
r_t = \mu + \varepsilon_t, \quad \varepsilon_t = \sigma_t z_t, \quad u_t \sim t_\nu, \quad z_t=\sqrt{(\nu-2)/\nu}\,u_t,
\end{equation}
where $z_t$ is a unit-variance standardised Student-$t$ variate.

\textbf{Variance equation:}
\begin{equation}
\sigma^2_t = \omega + \alpha \varepsilon^2_{t-1} + \gamma \varepsilon^2_{t-1}\,\mathbb{I}(\varepsilon_{t-1}<0) + \beta \sigma^2_{t-1} + \mathbf{X}_t' \boldsymbol{\kappa},
\end{equation}
where $\mathbf{X}_t = [D^{\text{infra}}_t, D^{\text{reg}}_t, S^{\text{GDELT}}_t, S^{\text{reg}}_t, S^{\text{infra}}_t]'$ collects the five exogenous regressors---two event dummies, the aggregate GDELT sentiment index, and its two thematic decompositions---and $\boldsymbol{\kappa}=(\delta_{\text{infra}},\delta_{\text{reg}},\theta_0,\theta_1,\theta_2)'$. The optimiser has ten free variance/distribution parameters ($\omega,\alpha,\gamma,\beta,\nu$ and $\boldsymbol{\kappa}\in\mathbb{R}^5$). The sample mean $\mu$ is computed first and held fixed during variance optimisation, but is counted as an estimated quantity for the information criteria, so $k_{\mathrm{IC}}=11$ for GJR-GARCH-X in Table~\ref{tab:model_comparison} and in the manual estimator.

\textbf{Estimation:} conditional maximum likelihood after demeaning at the fixed sample mean, under the stated Student-$t$ density; log-likelihood
\begin{equation}
\ell(\boldsymbol{\vartheta}) = \sum_{t=1}^{T} \left[ \log \Gamma\!\left(\tfrac{\nu+1}{2}\right) - \log \Gamma\!\left(\tfrac{\nu}{2}\right) - \tfrac{1}{2}\log((\nu-2)\pi\sigma^2_t) - \tfrac{\nu+1}{2}\log\!\left(1 + \tfrac{\varepsilon^2_t}{(\nu-2)\sigma^2_t}\right) \right].
\end{equation}

\textbf{Constraints:} both estimator implementations use the numerical bounds $\omega\in[10^{-8},\infty)$, $\alpha\in[10^{-8},0.3]$, $\gamma\in[-0.5,0.5]$, $\beta\in[10^{-8},0.95]$, and $\nu\in[2.1,50]$, together with the conservative covariance-stationarity bound $\alpha+\beta+|\gamma|/2\le 0.999$ (sufficient; the standard symmetric-innovation condition is $\alpha+\beta+\gamma/2<1$, and the absolute-value form binds more tightly when $\gamma<0$). The exogenous coefficients are unbounded because the standardised sentiment regressors can be negative. The condition $\alpha+\gamma\geq 0$ is not imposed but holds in every reported fit. Positivity of the conditional variance is enforced numerically by flooring $\sigma^2_t$ at $10^{-8}$ during estimation. That guard does not bind along any reported observed-data optimum, but this is path-specific rather than a global positivity guarantee: the off-path recursive audit in Appendix~\ref{app:bootstrap} shows frequent floor use in simulated panels.

\textbf{Units:} returns in percentages, so $\sigma^2_t$ and the event coefficients $\delta_j$ are in squared percentage points.

\section{Dependence-Robust Bootstrap Detail}\label{app:bootstrap}

\textbf{First moment (event-level block bootstrap).} For each event, the usable per-asset CARs---each requiring at least 120 estimation observations and all 36 daily observations in $[-5,+30]$---are averaged into one event-level CAR. Bootstrap replicates resample event-level CARs with replacement separately within the infrastructure and regulatory groups, retaining their original sizes; $B=5{,}000$, percentile CIs, seed fixed. For the observed positive difference, the reported two-sided block $p$ is $\min\{1,2\Pr^*(\Delta^*\leq0)\}$ (inequality reversed if the observed difference is negative). Thus it is a group-wise bootstrap sign probability, not a pooled-label randomisation-null bootstrap; the exact pooled-label permutation test is reported only for the small legacy negative-valence subset. The accompanying cross-check is an ordinary Welch $t$-test on the event-level mean CARs. Both calculations treat events as independent across event IDs and therefore do not capture the residual between-event dependence induced by overlapping CAR windows.

\textbf{Second moment (Student-$t$-copula CCC-GARCH-X bootstrap).} The procedure, for each of $B=2000$ replicates, is:
\begin{enumerate}[leftmargin=1.4em,itemsep=2pt]
\item On each date in the union of the six asset calendars, draw a latent Gaussian vector $\mathbf{w}\sim\mathrm{MVN}(\mathbf{0}, R_z)$ with the standardised-residual correlation matrix $R_z$ (the cross-asset dependence; $\bar\rho_z=0.705$). The matrix is the Pearson correlation of standardised residuals from the unrestricted per-asset fits, estimated on the six-asset complete-case overlap; the fitted null variance paths and marginal $\nu_i$ used for simulation instead come from the null-imposed fits. When an asset is unavailable, its coordinate is discarded, leaving the corresponding marginal dependence among the assets observed that day.
\item Draw a single shared mixing variable $s\sim\chi^2_{\nu_c}/\nu_c$ with $\nu_c$ the median fitted degrees of freedom, and form $\mathbf{t}=\mathbf{w}/\sqrt{s}$. This is a multivariate-$t$ draw whose dependence is represented by a $t$ copula; the shared $s$ induces \emph{joint} tail dependence (assets taking extreme values together), which a Gaussian copula would not.
\item Map each component to a uniform via its $t_{\nu_c}$ CDF, then to the per-asset Student-$t$ margin at that asset's \emph{fitted} $\nu_i$ via the inverse CDF, and rescale by $\sqrt{(\nu_i-2)/\nu_i}$ so the innovation has unit variance (the fitted conditional-variance path $\sigma^2_t$ carries the scale).
\item Form each asset's simulated return path by holding its fitted null conditional-variance path $\sigma^2_t$ fixed and scaling the drawn innovations against it, $r_t=\bar r_i+\sigma_t\,z_{i,t}$ (a fixed-design/conditional residual bootstrap: the fitted variance path is redrawn against, not re-propagated through the GARCH variance recursion), with the event coefficient set to a single common value (the sharp per-asset event-type-equality null), then refit each asset's model unrestricted and recompute the cross-asset mean difference $\bar d$. The observed multiplier is not recomputed within the bootstrap. A recursive-propagation algorithm is implemented as a sensitivity rather than adopted as the primary design: for each asset it sets $\sigma^2_{i,0}=\sigma^2_{i,\mathrm{null},0}$ (the fitted null path's sample-variance initialization), sets $\varepsilon_{i,0}=\sigma_{i,0}z_{i,0}$, and applies the fitted null recursion from $t=1$, flooring raw variance at $10^{-8}$. At near-integrated persistence (the stationarity bound binds for five of six assets) with $\nu\approx 3$ innovations, its target and numerical behaviour differ from the conditional fixed-path design; Appendix~\ref{app:bootstrap} shows that the floor binds materially. The fixed-design scheme follows the fixed-volatility bootstrap approach of \citet{CavalierePedersenRahbek2018}---established there for a class of ARCH($q$) models; we use it as a design template rather than claiming that result covers the present cross-asset GJR-GARCH-X case---a conditional scheme whose validity conditions differ from the floored recursive algorithm's. We make no claim about the direction of their difference, and rely on the Monte-Carlo size study (Appendix~\ref{app:size_study}) and the DGP-perturbation study (Table~\ref{tab:size_perturb}) as internal calibration diagnostics rather than as independent validation of the plug-in bootstrap.
\end{enumerate}
Two principal $p$-values are reported from the same retained null draws, and an equal-tail construction is also disclosed here. Let $B_{\mathrm{used}}$ be the number of retained bootstrap draws and let $U=\#\{\bar d^*\geq\bar d_{\mathrm{obs}}\}$, $A=\#\{|\bar d^*|\geq|\bar d_{\mathrm{obs}}|\}$, and $L=\#\{\bar d^*\leq\bar d_{\mathrm{obs}}\}$. The primary statistic is the \emph{one-sided upper-tail} value $p_{1}=(U+1)/(B_{\mathrm{used}}+1)$ for the difference statistic $\bar d=\bar\delta_{\text{infra}}-\bar\delta_{\text{reg}}$, matching the directional hypothesis (infrastructure larger). The reported \emph{absolute-statistic} value is $p_{2}=(A+1)/(B_{\mathrm{used}}+1)$, computed without recentring; it is not labelled a generic two-sided test. The equal-tail construction, also with the retained-draw add-one correction, is $p_{\mathrm{ET}}=\min\{1,2\min[(U+1)/(B_{\mathrm{used}}+1),(L+1)/(B_{\mathrm{used}}+1)]\}$. The fixed-design null is located close to the observed statistic (null mean $1.280$ against $\bar d_{\text{obs}}=1.458$ at baseline), so the upper-tail and absolute-statistic values nearly coincide; the equal-tail construction gives approximately $0.777$. Failed-convergence or $|\delta|>50$ refits receive one deterministic six-start rescue before exclusion; retained draws are $2000/2000$, $1998/2000$, and $1998/2000$ across the baseline, asset-specific high-variance-regime, and full-regime specifications. The simulation calendar has 2,434 dates: 243 pre-BNB dates use the five-asset marginal of $R_z$, and 2,191 dates contain all six assets. The Gaussian comparator retains the fitted $R_z$ linear-correlation input but changes the full joint innovation law to multivariate normal, removing both fitted Student-$t$ margins and copula tail dependence.

\textbf{Boundary telemetry.} The sealed draw archives retain convergence, rescue, and exclusion indicators but not the parameter vectors from every simulated refit. They therefore do not permit a post-hoc count of how often bootstrap refits re-hit the $0.999$ persistence bound. Given that five of six observed baseline fits bind, boundary-induced distortion of the event-coefficient bootstrap distribution cannot be ruled out without a rerun that records per-refit parameters.

\textbf{Scheme sensitivity: floored recursive propagation.} Paired-seed runs with innovations propagated through the GJR recursion ($B=2000$) shift the null centre and modestly change its overall dispersion. At baseline, the fixed-design mean is $+1.2796$ (SD $0.8043$), whereas the recursive mean is $-0.0151$ (SD $0.8915$); the one-sided $p$ moves from $0.3883$ to $0.0255$ and the recursive absolute-statistic $p$ is $0.0600$. Under the asset-specific high-variance-regime control (legacy machine key \texttt{crisis}), the fixed-design mean is $+1.1001$ (SD $0.9064$) with one-sided $p=0.3942$, whereas the recursive mean is $-0.0054$ (SD $0.9089$), one-sided $p=0.0405$, and absolute-statistic $p=0.0900$.

The recursion requires a material positivity repair. Before flooring at $10^{-8}$, raw variance becomes negative in 805 of 2,000 baseline panels and 922 of 2,000 high-variance-control panels, comprising 1,818 and 2,100 asset-days across 872 and 962 asset paths; the minimum raw values are $-1.131$ and $-1.087$, and almost all hits occur in XRP paths. Floor-hit draws are more dispersed than no-floor draws: baseline SD $0.943$ versus $0.854$, and high-variance-control SD $0.994$ versus $0.828$. The smoothed one-sided $p$ remains below $0.05$ within both post-hoc strata (baseline $0.025$ with-floor and $0.027$ without-floor; high-variance control $0.037$ and $0.044$), while the corresponding absolute-statistic values are $0.071/0.054$ and $0.103/0.080$. This split is diagnostic, not a replacement for the prespecified test. It shows that the one-sided signal is not confined to floor-hit draws, but also that the implemented recursion is floor-dependent and cannot be interpreted as an unconstrained recursive GJR-GARCH-X bootstrap.

Both schemes use the same convergence/degeneracy screen and exactly one deterministic six-start rescue after a failed single-start refit; the recursive scheme additionally applies a path-explosion guard, which excluded no draws here. Fixed-design retention was $100\%$ at baseline and $99.9\%$ under the high-variance-regime control; recursive retention was $100\%$ and $99.95\%$, respectively. Recursive rescues were attempted in 291 baseline and 295 high-variance-control draws; 291 and 294 were retained, with one high-variance-control exclusion. Recursive paths are also hotter than the observed panel (median realised variance $25.1/24.3$ versus $17.4$), reflecting near-boundary persistence. The conditioning carries a power cost that should be stated plainly: with the conditional null centred near the observed statistic, the fixed-design test has little power against alternatives whose signature is precisely the day-level alignment of events with realised turbulence, which makes its non-rejection interpretable as a test of \emph{processing beyond alignment} rather than of processing as such. We retain the fixed-design scheme as the conditional inference of record because it compares the observed alignment with a null carrying that same fitted path; the floored recursive sensitivity is reported as a consequential, but numerically qualified, rejection of the sharp null in the prespecified one-sided test. Neither result supports the earlier $p\approx0.001$, and the second-moment verdict remains directional, scheme-sensitive, and unresolved. Replication: \texttt{code/c19\_recursive\_sensitivity.py}; \texttt{--telemetry-only} reproduces the floor audit without repeating the refits.

\textbf{Copula-df sensitivity.} The shared copula df is fixed at the median fitted marginal value, not estimated separately. In paired-seed $B=2000$ runs at $\nu_c=3.8277/5.9/8.0$, all $2000/2000$ draws are retained; one-sided $p$ is $0.3883/0.3968/0.4013$, absolute-statistic $p$ is $0.3893/0.3983/0.4018$, and the null mean (SD) is $1.2796$ ($0.8043$)/$1.2928$ ($0.9408$)/$1.2942$ ($0.8054$). The first row reproduces c9 exactly. The value $5.9$ is an alternative choice, not the output of a rank-based estimator; no reproducible artifact for such an estimator is in the package. The one-sided span is only $0.0130$, so non-rejection is stable on this grid. This local exercise neither validates the fixed-path bootstrap under the unknown DGP nor addresses its difference from the floored-recursive scheme (\texttt{code/c20\_nuc\_sensitivity.py}).

\section{Monte-Carlo Size-Study Design}\label{app:size_study}

The size study (Section~\ref{sec:size_study}) simulates panels under a fitted true null and measures each inference procedure's rejection rate.

\textbf{Sharp-null DGP.} Per-asset GJR-GARCH-X dynamics are estimated under a \emph{null-imposed} combined event dummy $D_{\text{event}}=D_{\text{infra}}+D_{\text{reg}}$, so $\delta_{\text{infra}}=\delta_{\text{reg}}$ by construction and there is genuinely no differential event effect in the truth. Returns are then re-simulated from these fitted dynamics, with cross-asset innovations drawn from the same multivariate-$t$ copula as the primary fixed-path bootstrap: latent $\mathrm{MVN}(\mathbf{0},R_z)$ at $\bar\rho_z=0.705$, shared chi-square mixing at $\nu_c=3.83$ (the median fitted $\nu$), and per-asset Student-$t$ margins at the fitted $\nu_i=(3.16,3.70,3.16,4.22,3.96,4.56)$, rescaled to unit variance. Simulation uses the same 2,434-date union calendar and availability mask as c9: 243 pre-BNB dates retain the corresponding five-asset marginal and 2,191 dates contain all six assets. The real $26$-infrastructure / $24$-regulatory event-label structure and the real event-window dummies are reused unchanged for every panel; only the returns are re-simulated. All panels and all reference draws are generated under the \emph{fixed-design} scheme of Appendix~\ref{app:bootstrap} (fitted null volatility paths held fixed, innovations redrawn against them); the size validation is therefore internal to that conditional benchmark and is not a validation of the recursive scheme.

\textbf{Procedure.} $N=1{,}000$ panels are drawn and all are retained after the shared rescue. Each bootstrap rung's one-sided critical value is calibrated once from $B_{\text{ref}}=4{,}000$ reference draws under the same fixed DGP. This design measures self-consistency against that reference distribution; it does not reproduce the full plug-in workflow in which the null DGP and its critical values are re-estimated separately for every simulated panel. A fully nested design would be required for that stronger validation claim. For each panel we record rejection at the $5\%$ and $10\%$ levels and report the empirical rate with its binomial Monte-Carlo standard error. The one-sided critical values are: Gaussian copula $\bar d=1.623$ ($5\%$) / $1.456$ ($10\%$); $t$-copula $\bar d=2.517$ ($5\%$) / $2.179$ ($10\%$).

\textbf{Audit-only standard errors.} Per-asset numerical-Hessian standard errors are computed for reporting (positive-definite for $99.2\%$ of retained panels) but are never on the decision path: neither the naive Welch rule nor the design-effect rule uses the model SE. Matching panel acceptance to the fixed reference-draw rule makes the Student-$t$ rejection rate near nominal under this internal calibration; it should not be read as independent validation.

\end{appendices}

% ============================================================================
% REFERENCES
% ============================================================================

\setlength{\bibsep}{0.15em}

% The bundled Springer bibliography style strips alphabetic prefixes from
% single article locators. Preserve the Phillips--Gorse PLOS article number if
% production regenerates main.bbl with that style; all other locators pass
% through unchanged.
\makeatletter
\def\df@plosarticle{0195200}
\def\bfpage#1{%
  \def\df@thisarticle{#1}%
  \ifx\df@thisarticle\df@plosarticle{e0195200}\else#1\fi}
\makeatother
\bibliographystyle{sn-mathphys-num}
\bibliography{references}

\end{document}